\documentclass[onecolumn]{aa} 
\usepackage{siunitx}
\usepackage{cancel}
\usepackage{array}

\usepackage[stable]{footmisc}
\usepackage{amsmath}
\usepackage{txfonts}
\usepackage{placeins}
\usepackage{natbib}
\usepackage{comment}

\bibpunct{(}{)}{;}{a}{}{,}
\usepackage{graphicx}
\usepackage{subfig}
\usepackage{epstopdf}
\usepackage{ifthen}
\usepackage{mathtools}
\usepackage{fixltx2e}
\usepackage{url}
\usepackage[table,usenames,dvipsnames]{xcolor}
\usepackage{soul} 
\usepackage[export]{adjustbox}

\newcommand{\planck}{\Planck}
\providecommand{\Planck}{\textit{Planck}}
\providecommand{\planck}{\Planck}
\newcommand{\WMAP}{WMAP}
\def\wmap{\WMAP}
\providecommand{\core}{\textit{COrE}}

\newcommand{\Nside}{\ensuremath{N_{\mathrm{side}}}}

\providecommand{\healpix}{\texttt{HEALPix}}
\def\WMAP{{WMAP}}

\newcommand{\HW}{}                  

\begin{document}

   \title{Systematic effects induced by Half Wave Plate precession into Cosmic Microwave Background polarization measurements}
   
   \author{G. D'Alessandro,\inst{1,2}
          L. Mele,\inst{1,2}
          F. Columbro,\inst{1,2}
          L. Pagano,\inst{3,4,5,6}
          F. Piacentini,\inst{1,2}        
          P. de Bernardis,\inst{1,2}
          S. Masi,\inst{1,2}
          }

   \institute{$^1$Physics Department, Universit\`a di Roma 
	``Sapienza'', Ple.\ Aldo Moro 2, 00185, Rome, Italy\\
              $^2$ INFN -- Sezione di Roma1, Ple.\ Aldo Moro 2, 00185, Rome, Italy\\
              $^3$Dipartimento di Fisica e Scienze della Terra, Universit\`a degli Studi di Ferrara and INFN -- Sezione di Ferrara, Via Saragat 1, I-44100 Ferrara, Italy\\
              $^4$Institut d'Astrophysique Spatiale, CNRS, Univ. Paris-Sud, Universit\'{e} Paris-Saclay, B\^{a}t. 121, 91405 Orsay cedex, France\\
              $^5$Institut d'Astrophysique de Paris, CNRS, 98 bis Boulevard Arago, F-75014, Paris, France\\
			  $^6$LERMA, Sorbonne Universit\'{e}, Observatoire de Paris, Universit\'{e} PSL, \'{E}cole normale sup\'{e}rieure, CNRS, Paris, France\\
              \email{giuseppe.dalessandro@roma1.infn.it, lorenzo.mele@roma1.infn.it}
             }\label{inst1}

   \date{Received XXXXXXX XX, 2018; accepted XXXXXX XX, 2018}

\titlerunning{Precession in Half Wave Plate Polarimeter}
\authorrunning{G. D'Alessandro et al.}

\abstract {
The measure of the primordial B-mode signal in the Cosmic Microwave Background (CMB) represents the smoking gun of the cosmic inflation and it is the main goal of current experimental effort.
The most accessible method to measure polarization features of the CMB radiation is by means of a Stokes Polarimeter based on the rotation of an Half Wave Plate. }
{The current observational cosmology is starting to be limited by the presence of systematic effects. 
The Stokes polarimeter with a rotating Half Wave Plate (HWP) has the advantage of mitigating a long
list of potential systematics, by modulation of the linearly polarized component of the radiation, 
but the presence of the rotating HWP can by itself introduce new systematic effects, which 
must be under control, representing one of the most critical part in the design of a B-Modes experiment.
It is therefore mandatory to take into account all the systematic effects the instrumentation can induce. 
In this paper we present, simulate and analyse the spurious signal arising from the precession of a rotating HWP. }
{We first find an analytical formula for the impact of the systematic effect induced by the HWP precession on the propagating radiation,
using the 3D generalization of the M\"uller formalism.
We then perform several numerical simulations, showing the effect induced on the Stokes parameters by this  systematic. We also derive and discuss the impact into B-modes measured by a satellite experiment.}
{We find the analytical formula for the Stokes parameters from a Stokes polarimeter where the HWP follows a precessional motion with an angle $\theta_0$. We show the result depending on the HWP inertia tensor, spinning speed and on $\theta_0$. The result of numerical simulations is reported as a simple timeline of the electric fields. Finally, assuming to observe all the sky with a satellite mission, we analyze the effect on B-modes measurements.}
{The effect is not negligible giving the current B-modes experiments sensitivity, therefore it is a systematic which needs to be carefully considered for future experiments.}


\maketitle
%

\section{Introduction}
\label{sec:introduction}
In 2014, the BICEP2 experiment, designed to measure the cosmic microwave background (CMB) polarization, claimed the first detection of primordial B-modes \citep{BICEP2_2014}, measuring the tensor-to-scalar ratio as $\emph{r}=0.2^{+0.07}_{-0.05}$. One year later a joint effort  involving \Planck\ and BICEP2 collaborations \citep{Planck_BICEP2} revised this bound publishing an upper limit of $\emph{r}<0.12$, obtained by removing the residual dust contamination from the BICEP2 maps. More recently BICEP2 and Keck Array experiments \citep{Array:2015xqh} further reduce this constraint down to $\emph{r}<0.07$ which represents the current strongest constraint to date on inflationary gravitational waves.

Further improve the constraint \HW{on} the tensor-to-scalar ratio represents a hard challenge for the current and future CMB experiments which must take into account, accurately, gravitational lensing and foreground removal \citep{errard} in addition to an excellent control of systematic effects \citep{Wallis:2016bja}. Concerning the control over systematics, some experiments are designed with the capability of self-calibrating and thus of removing some systematic effects \citep{qubic_2012}. \HW{Others experiments,} which do not have this feature, an accurate instrumental systematics prevision and laboratory calibrations are mandatory in order to separate systematic errors from scientific data \citep{Natoli:2017sqz,Inoue_Polarbear,common_dale,Columbro}.

A standard \HW{device} for polarization analysis is the Stokes Polarimeter,  composed by an Half Wave Plate (HWP) and a polarizer. The HWP \citep{Pisano:08} induces a phase shift on the linear polarization by birefringence and the polarizer selects one component; so by rotating the HWP it is possible to modulate the linearly polarized fraction of the incoming radiation, and to extract the Stokes components (T, Q, U), given a reference frame. 
The HWP can spin fast ($>1Hz$), see e.g. \citet{LSPE2012, Columbro2019, ebex2017, Thornton_2016} in order to modulate the signal at high frequency and it can also \HW{rotate} step by step ($<1Hz$) \citep{qubic_2012, spider2016} depending on the experiment scanning strategy. Systematic effects like T-P leakage, chromatic HWP behaviour, scanning strategy, are already evaluated in \cite{Hileman_syst,hwp_systematics1,hwp_systematics2,Salatino_simon}, and also measured by \cite{Kusaka_sys}. 

In this paper we analyse the systematic effect induced in observation of the CMB polarization by 
the precession of the HWP, in the specific case of a transmissive plate. We assume the HWP free 
from other systematic effects.

In Sections~\ref{sec:precessing_hwp_theory} we review the dynamic of a spinning cylindrical plate, and define the precession rate.  
In section~\ref{sec:stokes_polarimeter} we provide an analytic study of the effect induced by the HWP which spins and precesses, \HW{by using the 3D Jones formalism, for fully polarized radiation, and the 3D generalization of the M\"uller formalism, for partially polarized radiation}. In section~\ref{sec:phenomenology} by using the equations derived before, we show some results on electric field components produced with numerical simulations. Finally, in section~\ref{sec:simulations} we describe the effect induced on full-sky observation of the CMB, assuming a satellite mission with different scanning strategies. 

\section{Precessing Half Wave Plate Theory} 
\label{sec:precessing_hwp_theory}

In this section we \HW{present equations describing} a precessing body rotating along one symmetry axis. We introduce the main variables showing their evolution with time. We report here only the main equations, essential for the results shown in the following sections. \HW{All the details} of the computation are provided in Appendix~\ref{app:pre_teo}.   

\begin{figure}[ht]
\centering
{\includegraphics[scale=0.35]{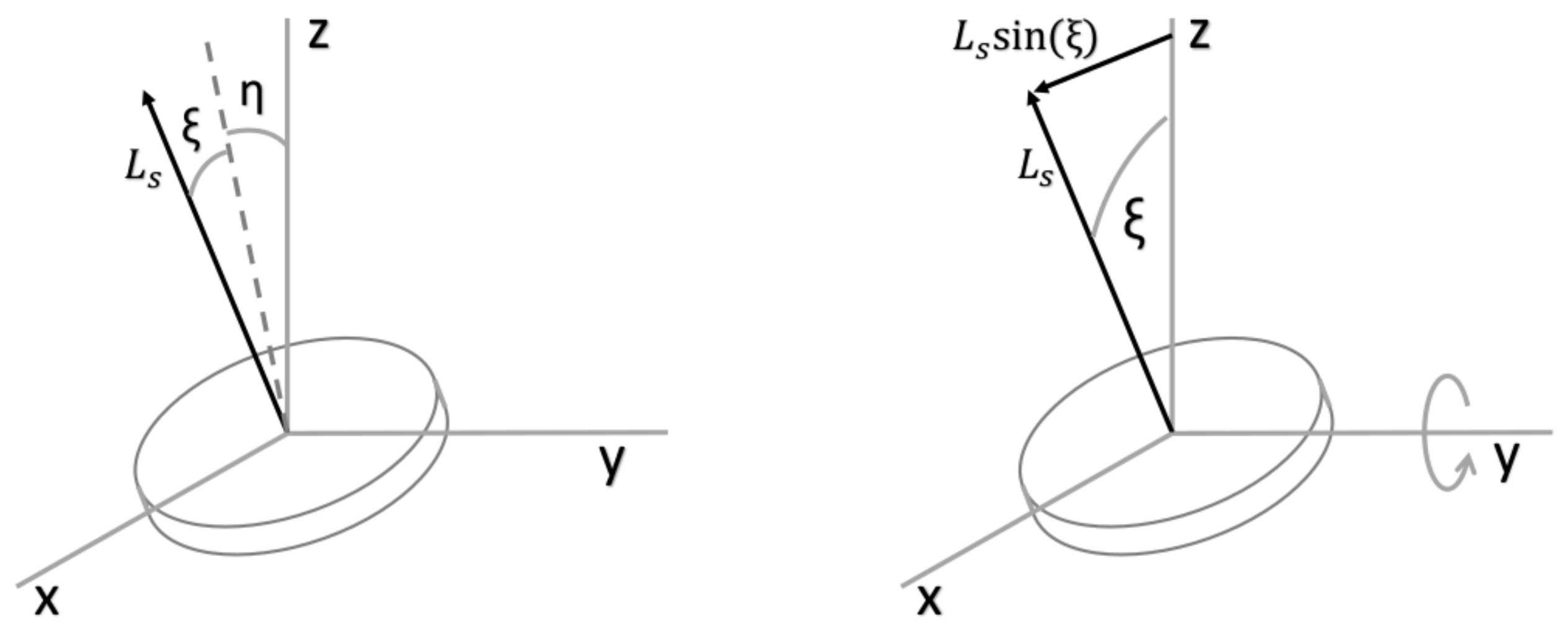}} \quad
  \caption{Illustration of rigid body precession where we can see the angular momentum dislocations respect to the $\hat{x}$ and $\hat{y}$ axes.} 
  \label{fig:body_precession}
\end{figure}

We approximate the HWP as a cylindrically symmetric rigid body, like a coin, and we define a reference system $\hat{x}$-$\hat{y}$-$\hat{z}$ with the $\hat{x}$-$\hat{y}$ plane coincident with the base of the cylinder, see \figurename \ref{fig:body_precession}.
We \HW{hypothesize} that the HWP has a large spin angular momentum $L_s = I_s \omega_s$ along the symmetry axis, where $I_s$ and $\omega_s$ are respectively the moment of inertia and the angular velocity. In the unperturbed case $L_s$ coincides with the $\hat{z}$ axis. 

We now introduce a small angular perturbation ($\theta_0$) of the spin momentum $L_s$, expanding it in its components along the $\hat{x}$ and $\hat{y}$ axes, $\eta$ and $\xi$. By introducing the precession rate $\gamma$, defined as the rotation velocity of $L_s$ around the $\hat{z}$ axis, $\eta$ and $\xi$ are simply defined as:
\begin{eqnarray}
\label{eq:theta_x}
\eta &=& \theta_0 \cos(\gamma t )\;, \\
\xi &=& \theta_0 \sin(\gamma t )\;.\label{eq:theta_y}
\end{eqnarray}

\noindent Those equations describe the torque-free precession of the spin axis that rotates in space with fixed angle $\theta_0$.

The precession rate is given by $\gamma = \omega_s I_s/I_\perp$ \citep{kleppner}, having defined $I_s$ and $I_\perp$ as the inertia momenta respectively parallel and perpendicular to the $\hat{z}$ axis.
Assuming the cylindrical plate as a thin disc we can approximate $I_s=2I_\perp$ and $\gamma=2\omega_s$, thus the disc wobbles twice as fast as it spins.

\noindent The apparent rate of a thin disc precession for an observer on the rigid body is given by:
\begin{equation}
\label{eq:gamma}
\gamma^\prime = \gamma -\omega_s=\omega_s\left(\frac{I_s-I_\perp}{I_\perp}\right) \sim \omega_s
\end{equation}

In a torque-free precession we can identify two different rotations accordingly to the reference frame we consider. In the fixed laboratory frame, the angular velocity vector rotates around the fixed $\hat{z}$ axis (where the angular momentum vector lies), tracing the so-called "space cone". In the reference frame integral with the rotating body, we can see both angular momentum and angular velocity vector describing a circle around the symmetry axis of the cylinder, tracing the so-called "body cone" with a precession rate $\gamma^\prime$.

\section{3D Generalization of Stokes Polarimeter}
\label{sec:stokes_polarimeter}

In literature a description of Stokes Polarimetry by using M\"uller and Jones matrices is already present, see e.g. \cite{Bryan_Muller_matrix}, \cite{ODea,Chuss:12}. In this section, and in Appendix \ref{app:stokes_pola} and \ref{app:stokes_pola2}, we \HW{derive} the analytic equation both for a traditional Stokes polarimeter and for a Stokes polarimeter where the HWP has a precessional motion.

\subsection{Jones formalism}

We now use an extended Jones formalism \citep{Jones1941} to retrieve a formula for Electromagnetic (EM) field intensity after the Stokes polarimeter. The Jones matrices used are defined in Appendix~\ref{app:stokes_pola} followed by explicit calculations.
\HW{The traditional Jones formalism can describe the state of fully polarized light} with a two-dimensional vector $i=(E_x, E_y$) and optical elements that change the state of input radiation with 2x2 matrices. We extend the Jones vector in three dimensions, $i=(E_x, E_y,0$), where the $\hat{z}$ axis is the propagation direction of the EM field, and the $\hat{x}$-$\hat{y}$ plane is perpendicular to $\hat{z}$, thus any optical element is described by 3x3 matrices \citep{Sheppard:11,Sheppard:14,Ortega-Quijano:13}.
Combining such matrices we get a general formula for the Jones vector for the on-axis detector of a \HW{precessing polarimeter}:

\begin{equation}
\label{eq:intensity_vector}
	i_{out}=J_{pol} \cdot J_{RotY}^{-1}(\xi) \cdot J_{RotX}^{-1}(\eta) \cdot J_{HWP}(\theta) \cdot J_{RotX}(\eta) \cdot J_{RotY}(\xi) \cdot i_{in}
\end{equation}

\noindent
For the ideal case with $\eta=\xi=0^\circ$, when no precession occurs, we get the outgoing intensity from an ideal polarimeter:
\begin{equation}
I = \frac{1}{2}\left[T+Q\cos(4\theta)+U\sin(4\theta)\right].
\label{eq:ideal_polarimeter}
\end{equation}
For the general case with $\eta$, $\xi \neq 0^\circ$ we find the intensity at the detector: 
\begin{equation}
\label{eq:intensity_general_symbol}
I= i_{out_x}^2 + i_{out_y}^2 +i_{out_z}^2  =\left( E_x g - E_y f\right)^{2}
\end{equation}

\noindent where we define the modulating functions $g(\xi,\theta)$ and 
$f(\eta,\xi,\theta)$: 
\begin{eqnarray}
\label{eq:g_f}
g&=&\sin^{2}{\left (\xi \right )} + \cos^{2}{\left (\xi \right )} \cos{\left (2 \theta \right )}\nonumber ,\\
f&=&2 \sin{\left (\theta \right )} \cos{\left (\xi \right )} \left(\sin{\left (\eta \right )} \sin{\left (\xi \right )} \sin{\left (\theta \right )} -  \cos{\left (\eta \right )} \cos{\left (\theta \right )} \right)
\end{eqnarray}

\noindent We can write the outgoing intensity through the Stokes parameters as follows:
\begin{equation}
\label{eq:intensity_general}
I = \frac{1}{2} \left( g^2 + f^2  \right) T + \frac{1}{2} \left( g^2 - f^2  \right) Q + g f U
\end{equation}

\noindent
\HW{valid only if $T^2 = Q^2 + U^2$.} The last equation gives the intensity for the on-axis detector of a Stokes polarimeter with its modulating element describing a torque-free precession. It is not merely a function of the HWP orientation about the $\hat{z}$ axis, it also depends on the displacements from the $\hat{x}$ and $\hat{y}$ axes due to the precession.

\subsection{M\"uller formalism}
We \HW{now} use a 3D extended M\"uller formalism \citep{ Sheppard:16, Samim:article} to retrieve a formula for EM field intensity outgoing from the Stokes polarimeter. 
\HW{The M\"uller formalism (fully defined in Appendix~\ref{app:stokes_pola2}) is required for the case of CMB experiments since we want to propagate partially polarized radiation through a polarimeter in order to extract the information about its polarization state.
Starting from 3D extended Jones matrices (Appendix \ref{app:stokes_pola}), the $9x9$ M\"uller matrix corresponding to each optical element can be easily obtained from Eq.~\ref{eq:Muller_to_Jones}: }
\begin{equation}
\label{eq:Muller_to_Jones}
M_{ij}= tr(\sigma_i \cdot J \cdot \sigma_j \cdot J^{\dagger})
\end{equation}
\HW{where $\sigma_n$ ($n=[0,...,8]$) are the trace-normalized Gell-Mann matrices \ref{eq:gellman}. The Stokes polarimeter M\"uller matrix becomes:}
\begin{equation}
M_{\textit{SP}_\textit{wob}} (\eta, \xi, \theta) = M_{pol} \cdot M_{RotY}^{-1}(\xi) \cdot M_{RotX}^{-1}(\eta) \cdot M_{HWP}(\theta) \cdot M_{RotX}(\eta) \cdot M_{RotY}(\xi)
\end{equation}
\HW{So, the output Stokes vector is: }
\begin{equation}
s_{out}=M_{\textit{SP}_\textit{wob}} (\eta, \xi, \theta) \cdot s_{in}  
\end{equation}
\HW{Each component of the Stokes vector is defined in Eq.~\ref{eq:stokesvector} and assuming the field entering the polarimeter has $E_z=0$, $$s_{in}^T = (\Delta_0, \Delta_1, 0, 0, \Delta_4, 0, 0, 0, \frac{1}{\sqrt{2}}\Delta_0)$$}
\HW{Thanks to Eq.~\ref{eq:stokes_relation} which describes the relation between the 3D Stokes components and the classic definition of Stokes parameters we can find the outgoing intensity:}
\begin{equation}
    I=\frac{1}{\sqrt{6}}\left(\Delta_0+3\left(m_{01}\Delta_1 + m_{04}\Delta_4 + m_{08}\Delta_8\right)   \right) =
    \left(\frac{1}{3}+\frac{\sqrt{2}}{2}m_{08}\right)T+\frac{\sqrt{6}}{2}m_{01}U + \frac{\sqrt{6}}{2}m_{04}Q
    \label{eq:intensity_woobled_Muller}
\end{equation}
\HW{where $m_{ij}$ are the components of the Stokes polarimeter M\"uller matrix and are function of to the wobbling angles $\eta, \xi$.  
Note that if $\eta = \xi = 0$ the Eq.~\ref{eq:intensity_woobled_Muller} gives the well-known formula of the Stokes polarimeter (Eq.~\ref{eq:ideal_polarimeter}).}

\section{Phenomenology}
\label{sec:phenomenology}
\begin{figure}[ht]
\centering
  (a){\includegraphics[scale=0.33]{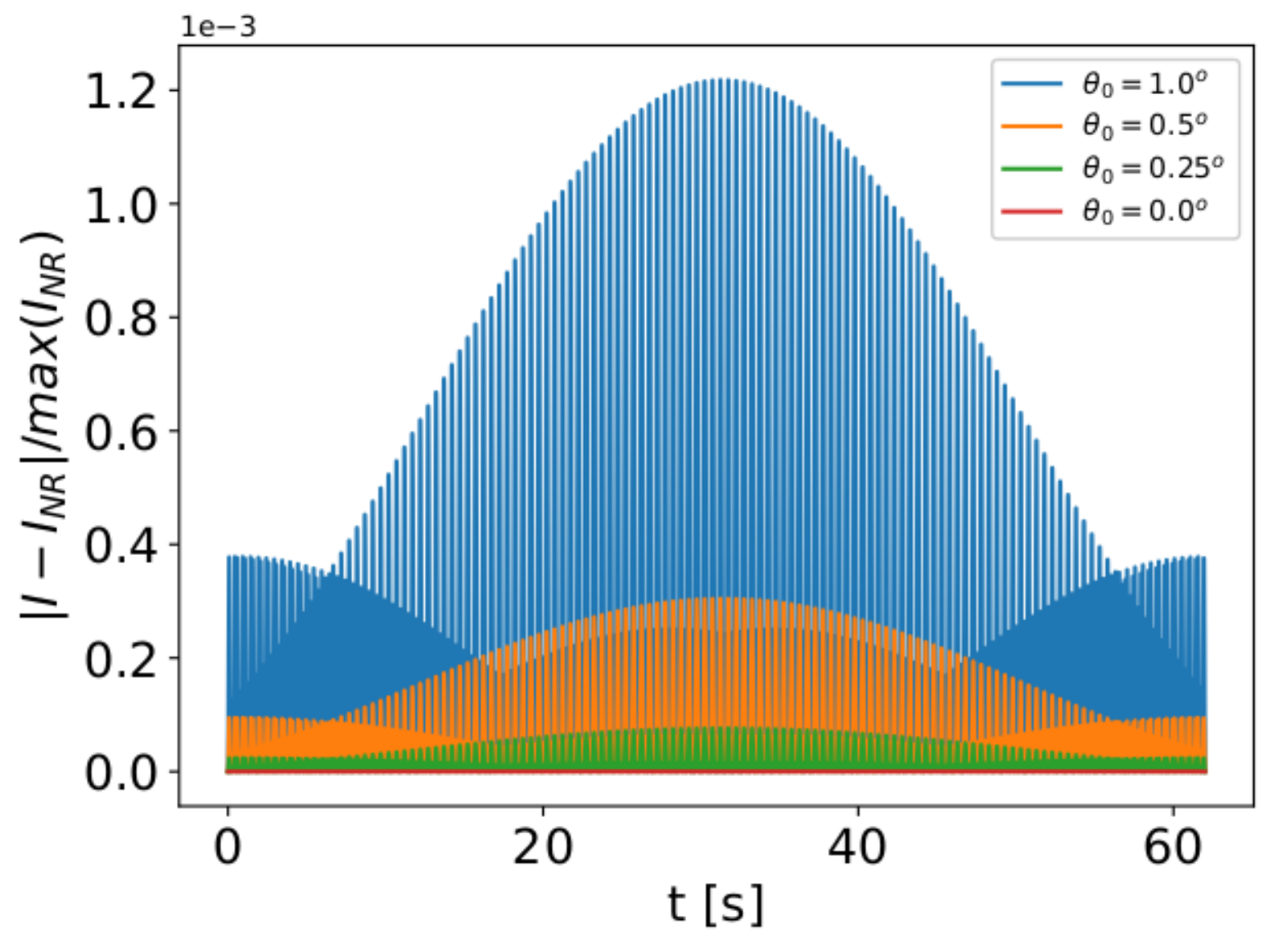}}  \quad
  (b){\includegraphics[scale=0.33]{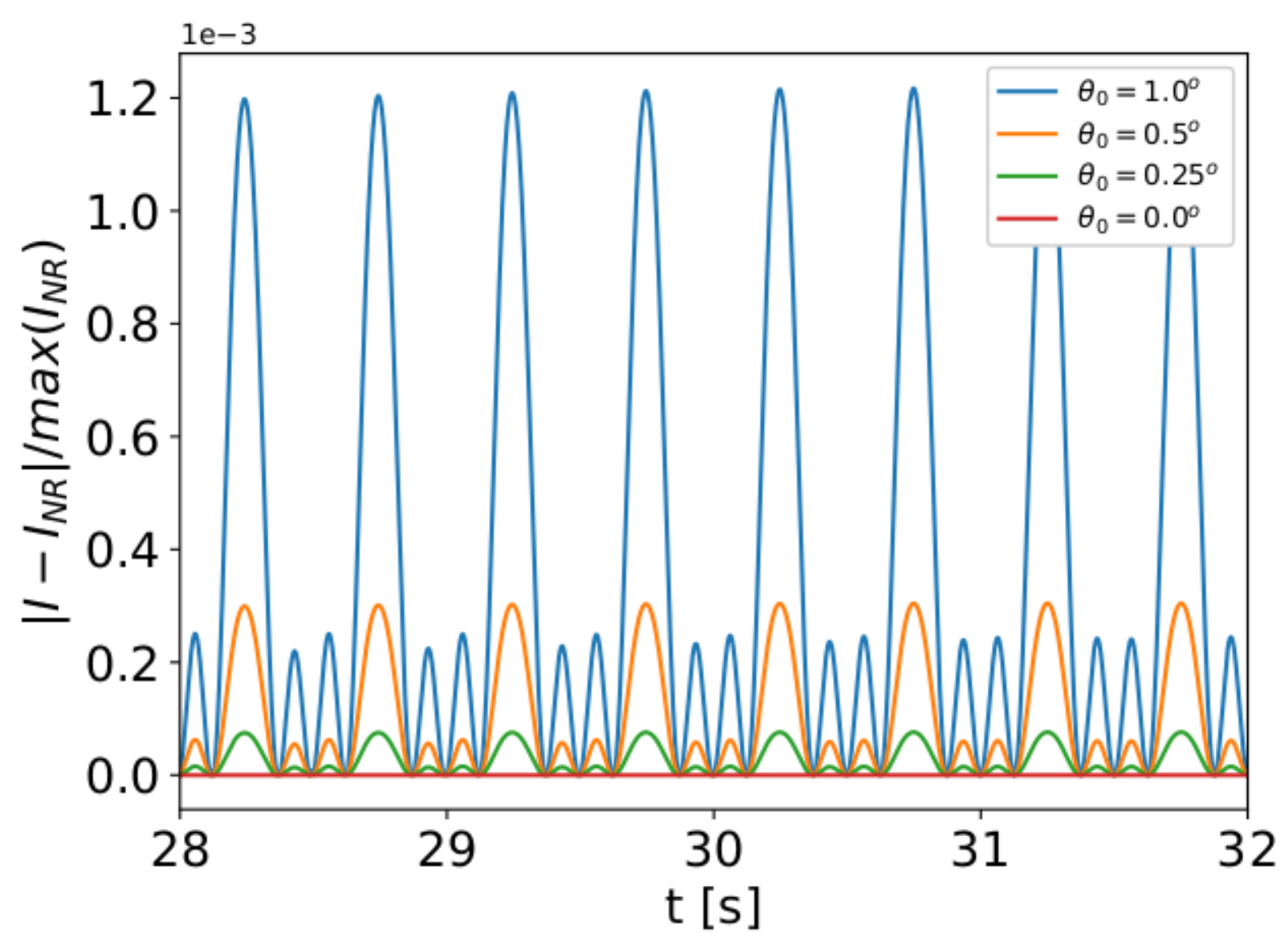}}  \quad
  (c){\includegraphics[scale=0.33]{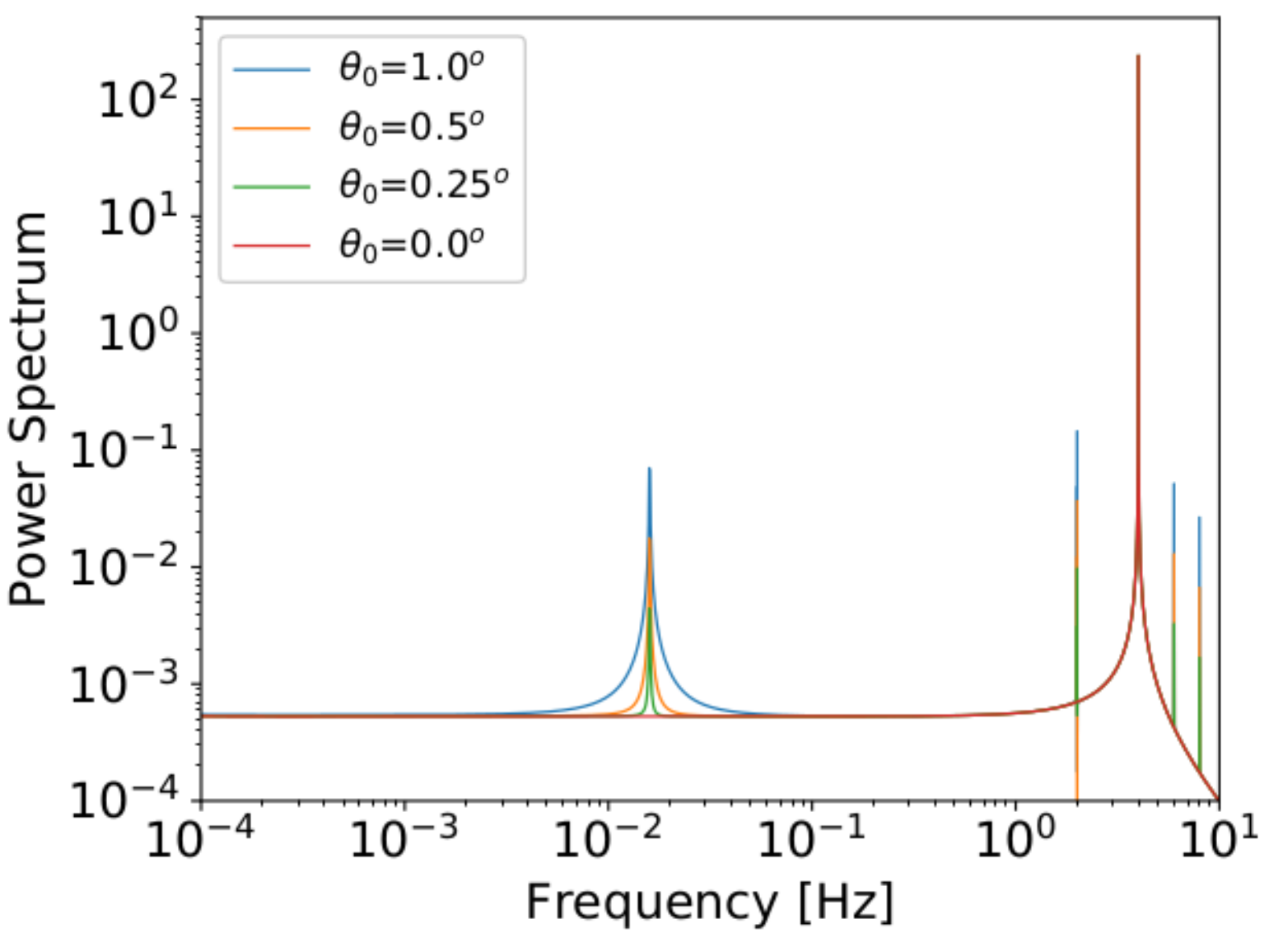}} \quad
  \caption{\HW{(a) Fractional residuals between the total intensities in a Stokes polarimeter with and without the HWP precession. We fix the ratio $I_{\perp}/I_s=0.502$ and consider different HWP precession angles $\theta_0=[1^\circ, 0.5^\circ, 0.25^\circ]$, normalizing the difference respect to the maximum intensity of an ideal Stokes polarimeter. (b) Fractional residuals, as before, but in a shorter time interval to highlight the beats induced by precession. (c) Power spectra with peaks at multiple and sub-multiple of the HWP spinning frequency with an amplitude related to $\theta_0$.}}
  \label{fig:intensity_diff}
\end{figure}

Starting from Eq.~\ref{eq:intensity_general}, we built a simulation to show the effects of the precession for a specific case. We set the spinning frequency at 1Hz and the ratio $I_{\perp}/I_s=0.502$.
The Jones vector used as input field in the simulation is $i_{in}=(1,0,0)$, corresponding to (1,1,0,0) Stokes vector.
Typical outputs of the simulations are reported in \figurename~\ref{fig:intensity_diff}.
We show the fractional residuals, defined as the difference between the Intensity from a precessing HWP and from the ideal non-precessing case, normalized to the maximum Intensity of the ideal polarimeter. 

The timelines reported in \figurename~\ref{fig:intensity_diff}(a) show the fractional residual for different amplitudes of the precession angle $\theta_0$. 
These timelines clearly show several beats with an amplitude depending on the precession angle $\theta_0$ in a non-linear way.

In \figurename~\ref{fig:intensity_diff}(c) we report the power spectra of timelines, for the ideal case (red), and for the precessing cases. The ideal case presents a single line at 4Hz, as expected from a wave-plate spinning at 1Hz. The beats from a precessing HWP produce spurious peaks at multiple and sub-multiple of the spin frequency.
The most noteworthy peak in the spectra is the one at lower frequencies as it is the result of the very slow modulation imposed by the precession, that in \figurename~\ref{fig:intensity_diff}(a) has a period of about 60s, implying a peak at 0.016Hz in the spectrum.
In the next section we illustrate how these beats depend on the geometrical parameters of the plate. 

\subsection{Spinning Speed and ratio $I_{\perp}/I_s$ effects} 
\label{sec:spinning}

For a cylindrical HWP, including its support, with mass $m$, thickness $h$ and radius $R$, the components of the moment of inertia respect to the principal axes are:

\begin{eqnarray}
I_s&=&\frac{1}{2}m R^2;\\ \label{eq:inertia_par} 
I_{\perp}&=&\frac{1}{12}m \left(3R^2 + h^2\right). \label{eq:inertia_perp}
\end{eqnarray}

where we are assuming a diagonal inertia tensor: 
\begin{equation}
\label{eq:inertia_matrix}
	\bf{I}=
	\begin{bmatrix}
	I_{\perp} & 0 & 0\\
	0 & I_{\perp} & 0\\
	0 & 0 & I_s\\
	\end{bmatrix}.
\end{equation}
The frequency for the precessional motion is directly linked to the HWP regular spin frequency $f_s$ and to the ratio $I_{\perp}/I_s$ (Eq.~\ref{eq:freq}). 

\begin{figure}[ht]
\centering
{\includegraphics[scale=0.55]{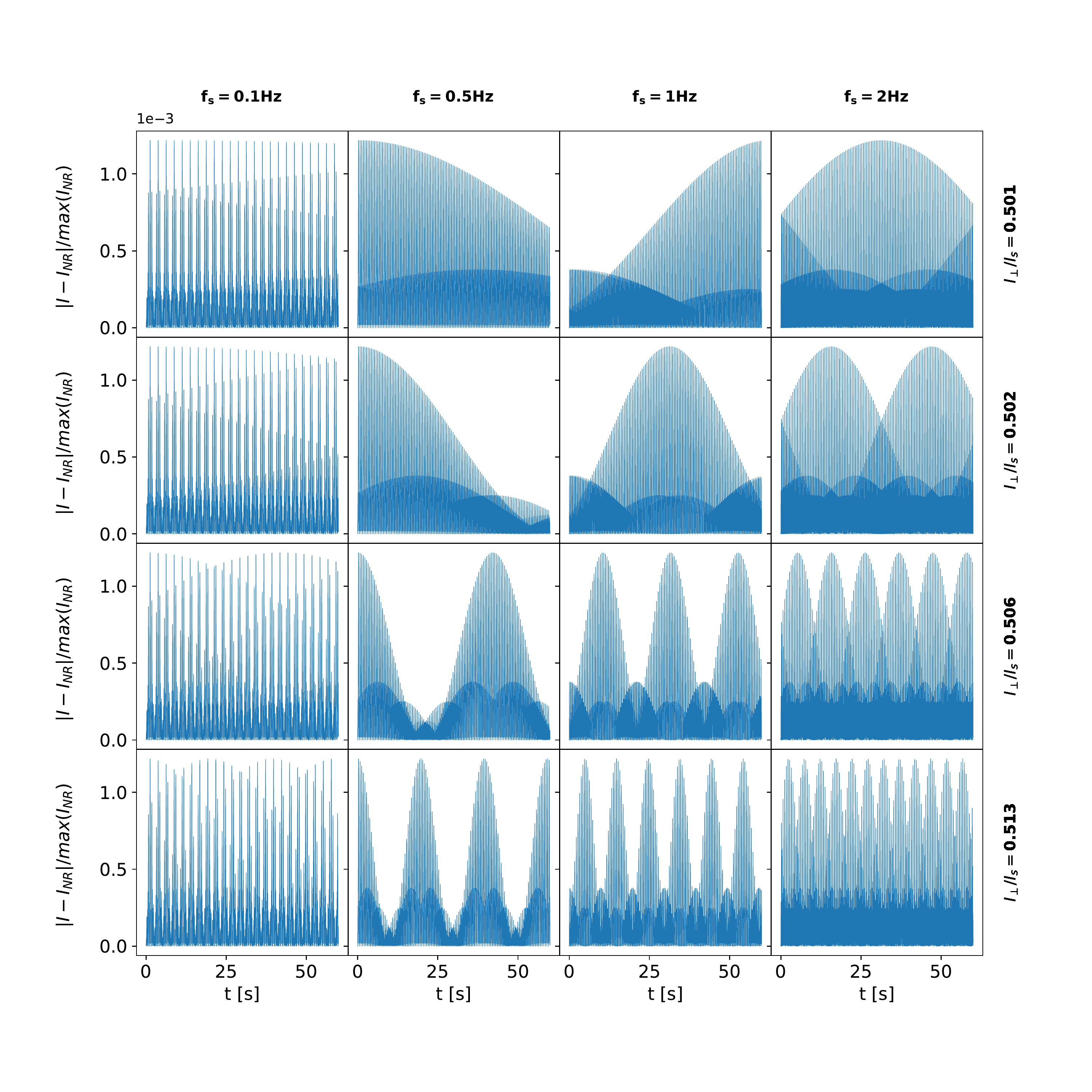}} \quad
  \caption{\HW{Fractional residual for different ratios $I_{\perp}/I_s$ and different HWP spinning frequencies. Each column shows a different $f_s$ value while each row a different $I_{\perp}/I_s$ ratio. Note that a thinner HWP, $I_{\perp}/I_s \rightarrow 0.500 $, has beats in the Intensity over an extended period while a thicker one has shorter beats. By looking the Figure from left to right it is clear how the spinning speed compress the beats. The maximum amplitude remains the same because it depends only on $\theta_0$ that is fixed to $1^\circ$ in this particular simulation.}} 
  \label{fig:intensity_plots}
\end{figure}

We can note that this ratio depends only on the cylinder height and radius as:
\begin{equation}
I_{\perp}/I_s=0.5+\frac{1}{6} \left( \frac{h}{R} \right)^2.
\end{equation}
We therefore explore different configurations as shown in \figurename~\ref{fig:intensity_plots} where we show the fractional residual with respect to the ideal case. 
We consider input radiation horizontally polarized, a precession angle $\theta_0 = 1^\circ$, spin frequencies $f_s=[ 0.1, 0.5, 1.0, 2.0 ] Hz$ and $I_{\perp}/I_s=[0.501, 0.502, 0.506, 0.513]$. 
As an example, these values for the $I_{\perp}/I_s$ ratio correspond to an HWP with mass $m=1kg$, radius $R=16cm$, and thickness $h\simeq [1.2, 1.7, 3.0, 4.5]~cm$; then $I_s$ is fixed to $0.0128$ $kg\cdot m^2$.
In practice the ratio of the components of the moment of inertia does not depend only on the plate thickness, but also on the structure of the HWP support.  

The \figurename~\ref{fig:intensity_plots} shows the dependence of the effect from the HWP spinning speed and the inertia tensor: the simulation shows that a thinner HWP, $I_{\perp}/I_s \rightarrow 0.500 $, has beats in the Intensity over an extended period while a thicker one has shorter beats. It is clear by looking the \figurename~\ref{fig:intensity_plots} from top to bottom. Anyway the value $0.5$ is not possible since it corresponds to a null thickness.

The effect of different spinning speeds is to compress the beats. This is clear by looking the \figurename~\ref{fig:intensity_plots} from left to right. 
The maximum amplitude remains the same because it depends only on $\theta_0$ that is fixed to $1^\circ$ in this particular simulation.

\begin{figure}[ht]
\centering
  {\includegraphics[scale=0.5]{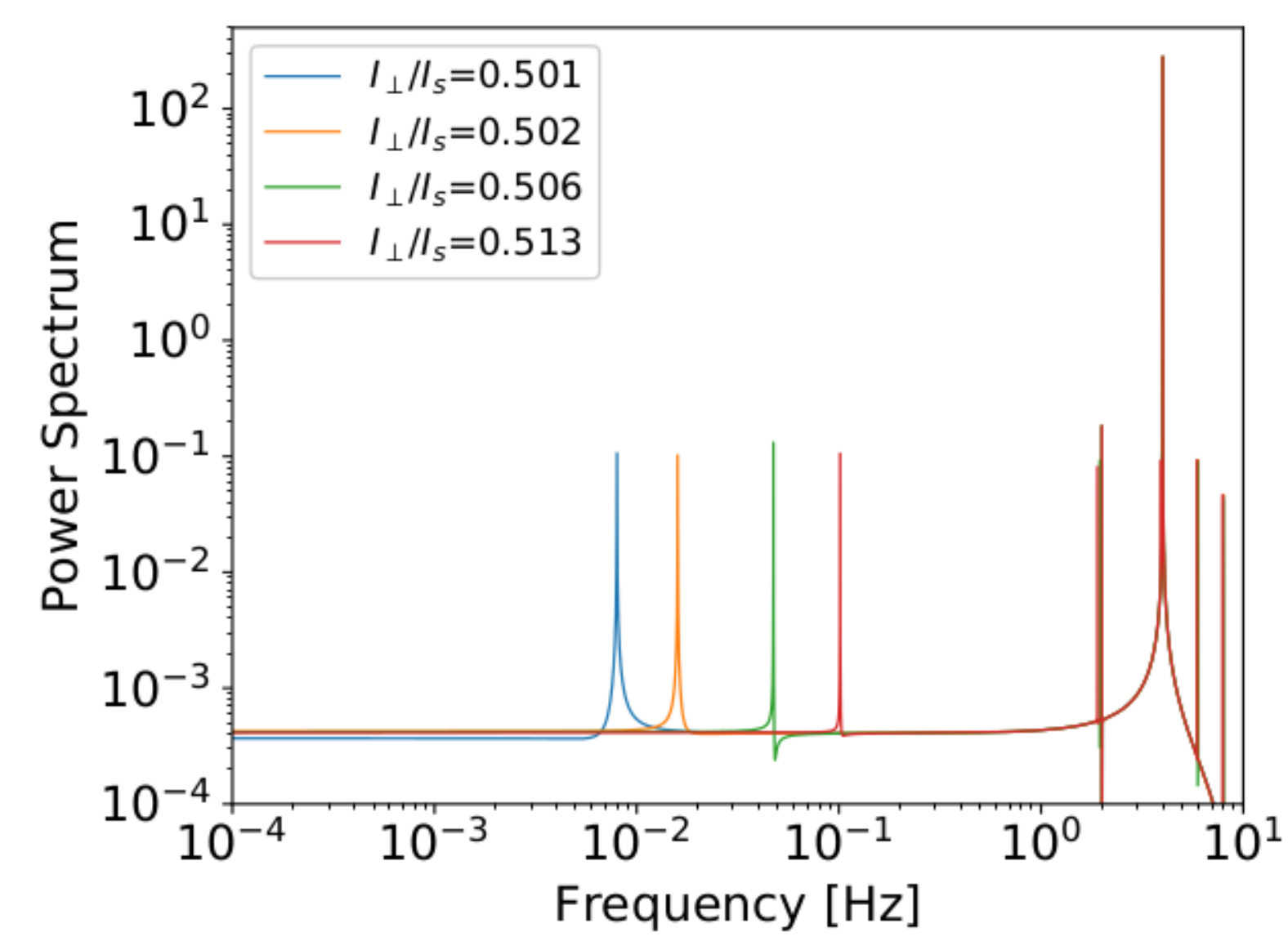}}  \quad
  \caption{\HW{Power spectra extracted from timelines showing the shifting of the spurious peak at low frequencies.
The simulation considers input light horizontally polarized, a precession angle $\theta_0 = 1^\circ$ and $I_{\perp}/I_s=[0.501, 0.502, 0.506, 0.513]$.}}
	\label{fig:power_freq}
\end{figure}

In Fig.~\ref{fig:power_freq} we report the power spectra of the timelines for different ratios $I_{\perp}/I_s$. The spectra exhibit the effect discussed above, showing that the beats frequency moves to lower values as the ratio $I_{\perp}/I_s$ is reduced towards the minimum value of 0.5 (for $I_{\perp}/I_s=0.513$ the period of the beats is 10s, corresponding to a peak in the spectrum at 0.1Hz).

\subsection{Minor effect}
As reported e.g. in \cite{Honggang}, a tilted HWP also changes its transmission properties due to the different path of radiation inside the plate with respect to the case of normal incidence. 
The resulting effect is a variation of retardance, which can be as high as $40\%$ in the case discussed in \cite{Honggang}, for a source at 347 nm, with 5 degrees tilt. 
This effect is multiplicative with the ratio wavelength/thickness, which is much higher in the case of plates used in the optical bandwidth with respect to the case of the millimeter wavelength. In instrumentation devoted to observations of the CMB polarisation, the plate thickness is of the same order of magnitude as the wavelength: 3.1mm POLARBEAR-2 \citep{PolarBear}, 3.05mm SPIDER \citep{spider_t}, 1.62mm EBEX \citep{ebex2017}, 3.2mm QUBIC \citep{Qubic_t}, 3mm LSPE-SWIPE \citep{LSPE2012}, 3.05mm ABS \citep{ABS}.
The impact of the effect for millimeter astronomy can be examined through the variation between the input Stokes vector and the output one, and it is estimated to be orders of magnitude smaller than the effect due to electromagnetic field projection analyzed in the rest of this paper.

\section{Full-Sky Simulations}
\label{sec:simulations}

In order to test the impact of the HWP precessional motion on CMB observations we build an algorithm able to generate a realistic satellite scanning strategy in presence of spinning HWP, producing data timelines. We complete this software with a map-making algorithm which collapses data timelines into maps. All simulations are noise-free, to better capture the impact of systematic effects. 

\subsection{Simulation pipeline}
\label{sec:simulator}
The scan simulator takes as inputs the details of a Satellite scanning strategy, three spin rates and two angles (see \cite{Das:2013nfa} for a detailed description of the geometrical configuration), namely:

\begin{itemize}
\item Earth revolution velocity $\omega_1$,
\item precession velocity $\omega_2$,
\item satellite spin $\omega_3$,
\item precession angle $\alpha$, i.e. the angle between the satellite spin axis and the sun-earth direction,
\item boresight angle $\beta$, i.e. the angle between the focal plane direction and the spin axis.
\end{itemize}

We simulate only a single detector placed at the centre of the focal plane illuminating a spinning HWP with  $f_s$ frequency. The systematic affecting the HWP is included in the data at the timeline level and a simple re-binning map-making is used to average all the samples in T,Q,U Stokes parameters maps \citep{Tegmark:1996qs}. In this paper we consider \planck\ \citep{Planck:2006aa}, \WMAP\ \citep{Bennett:2003ba}, \core\ \citep{Natoli:2017sqz} and LiteBIRD \citep{LiteBIRD:18} scanning strategies. The input parameters we employ for those scanning strategies are listed in Tab.~\ref{tab:scanning}. As sampling rate we use $60$ Hz.

\begin{table}[h!]
\centering
\begin{tabular}{lcccc} 
 Scanning & $\alpha\;[deg]$ & $\beta\;[deg]$ & $\omega_2 [deg/min]$  & $\omega_3 [deg/min]$\\[0.4ex]  
 \hline
\planck\ like & 7.5 & 85.0 & 0.00139 & 360.0\\
\wmap\ like & 22.5 & 70.5 & 6.0 & 167.0\\ 
\core\ like & 30.0 & 65.0 & 0.0625 & 180.0\\
LiteBIRD like & 45.0 & 50.0 & 3.8709 & 36.0 \\[0.4ex]
\end{tabular}
\caption{Parameters for the scanning strategies adopted in the simulation pipeline.}
\label{tab:scanning}
\end{table}

\subsubsection{Input map}
\label{sec:input}
The input sky map, used for full-sky simulations, contains: Solar Dipole and Galactic diffuse foregrounds in temperature and a CMB realization, both in temperature and polarization. We decide to include foregrounds only in temperature in order to highlight the temperature to polarization leakage. The input $\mathcal{C}_\ell$ used for the CMB realization is compatible with the best fit of \planck\ 2015 release \citep{Ade:2015xua} with no tensor perturbations.

The foreground field is generated from the Commander solution delivered with the \planck\ 2015 release \citep{Adam:2015tpy}. It includes the primary temperature emissions \citep{planck2014-ES}: synchrotron, free-free, spinning dust, CO and thermal dust  emission, without considering their polarization contribution.

Such map is modelled in order to highlight the temperature to polarization leakage induced by the HWP precession during the observations.
We set the resolution parameter of the input map at \healpix\ \citep{Gorski:2004by} resolution $\Nside=256$, comfortable enough respect to the Gaussian beam with FWHM=60 arcminutes. 
In order to evaluate the impact of parameters
chosen for the simulation, we have run a case with $\Nside=128$,
finding the same results in terms of angular power spectra
residual, except in the smaller scales, where the pixel 
size matters independently of the 
presence of systematic effects. 

\subsection{Maps and results}
\label{sec:maps_results}

\subsubsection{Output maps}
\label{sec:output}
\begin{figure}[ht]
\centering
  \includegraphics[scale=0.28,trim=1.4cm 0 .5cm 0,clip=True]{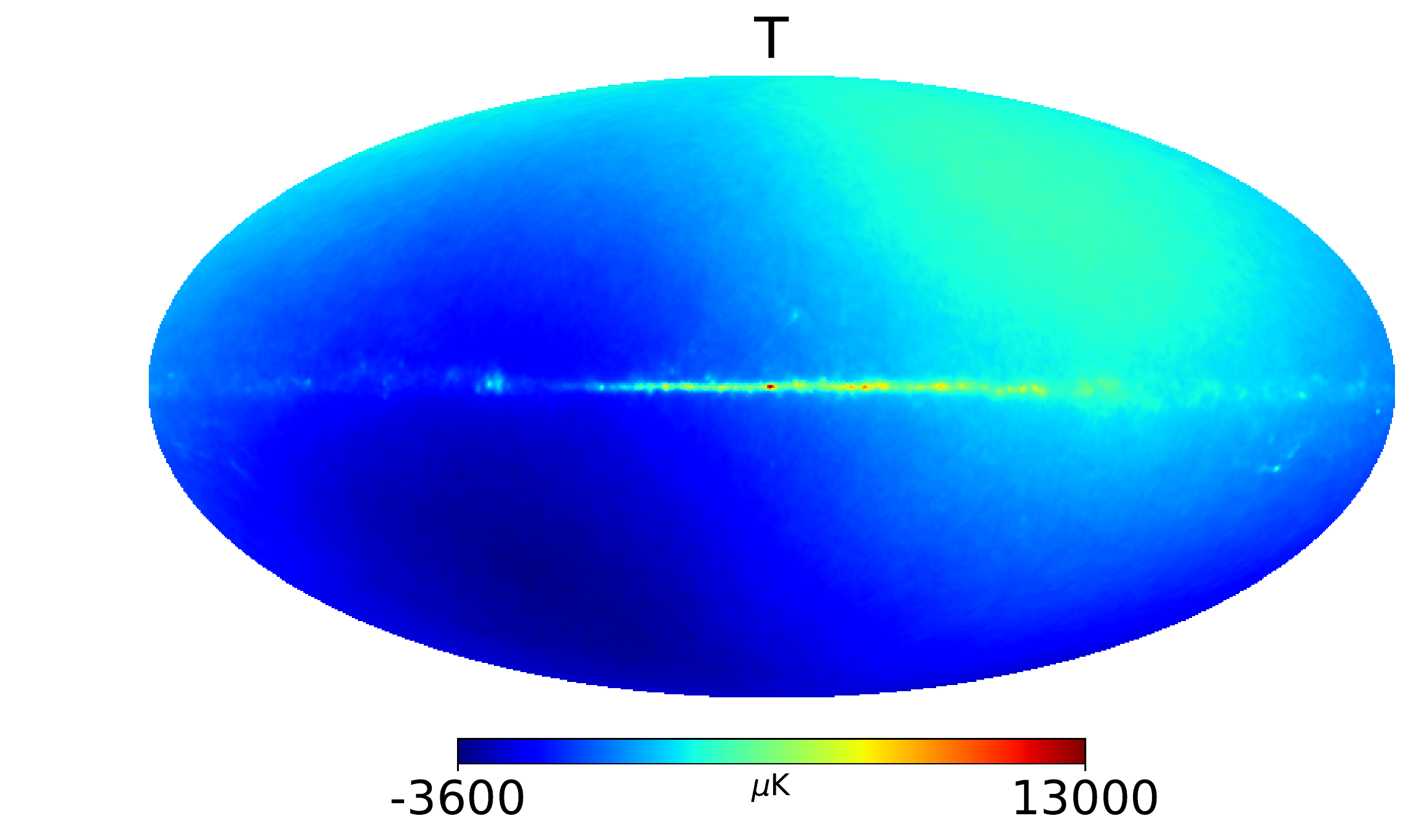} 
  \includegraphics[scale=0.28,trim=1.4cm 0 .5cm 0,clip=True]{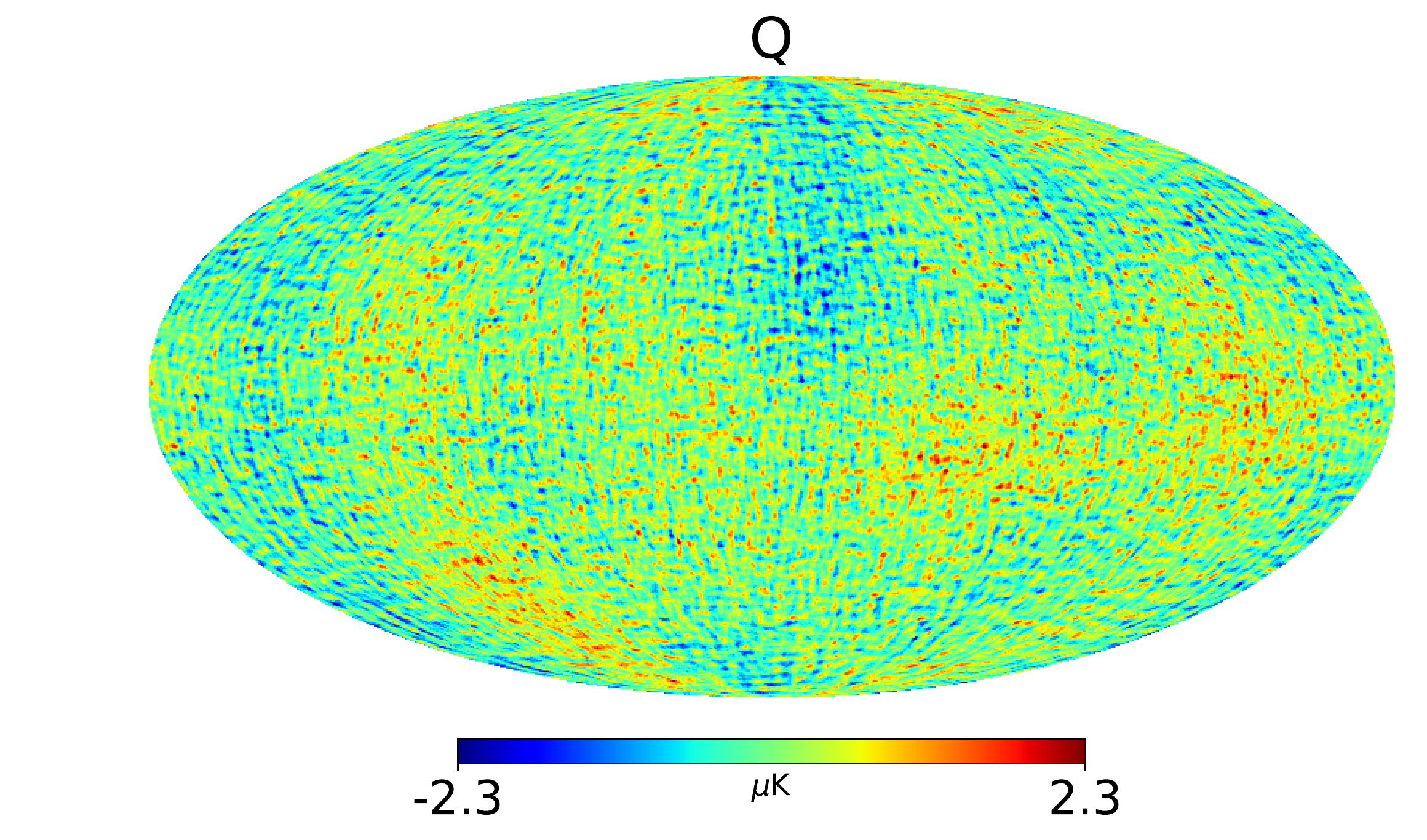}   
  \includegraphics[scale=0.28,trim=1.4cm 0 .5cm 0,clip=True]{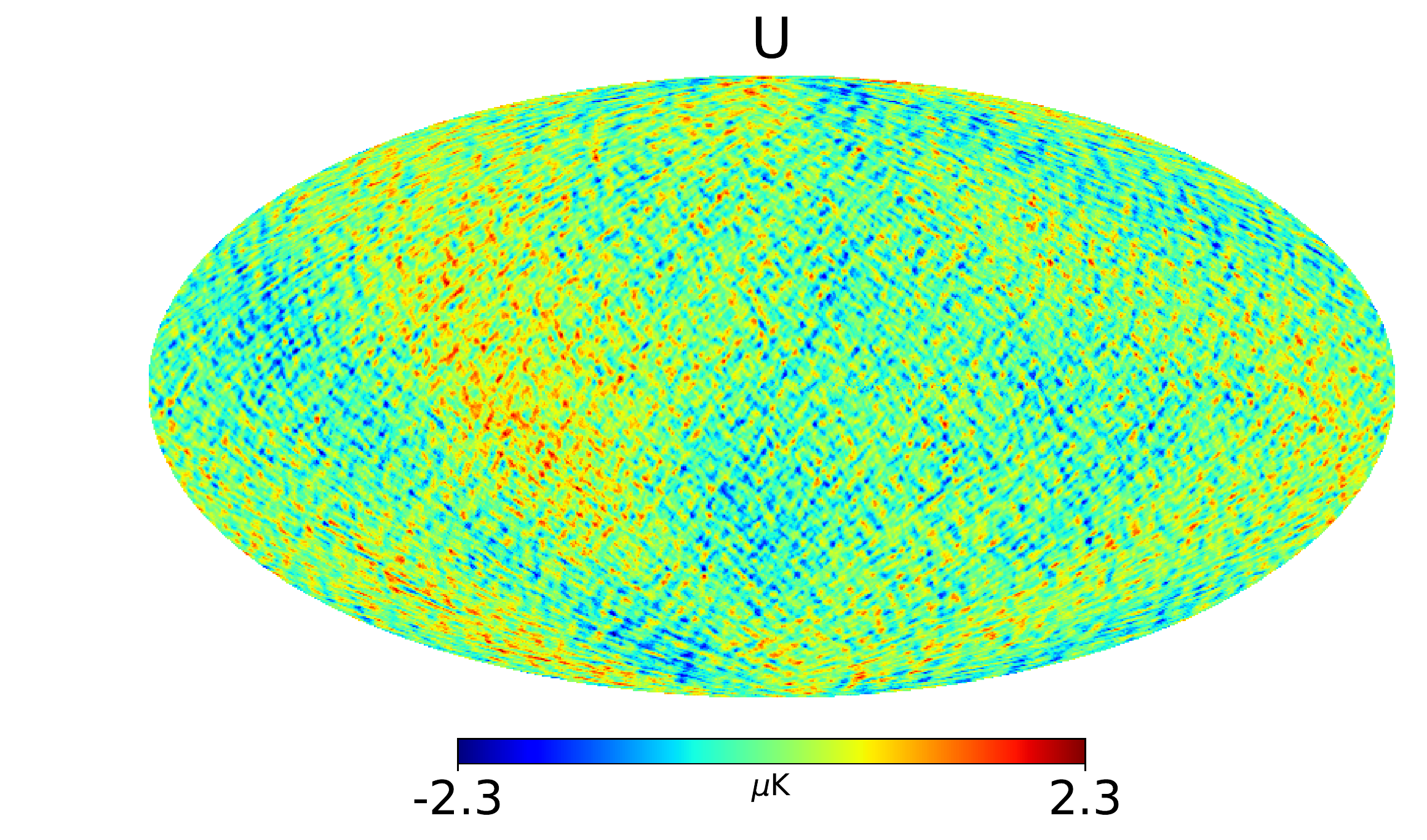}\\
  \includegraphics[scale=0.28,trim=1.4cm 0 .5cm 0,clip=True]{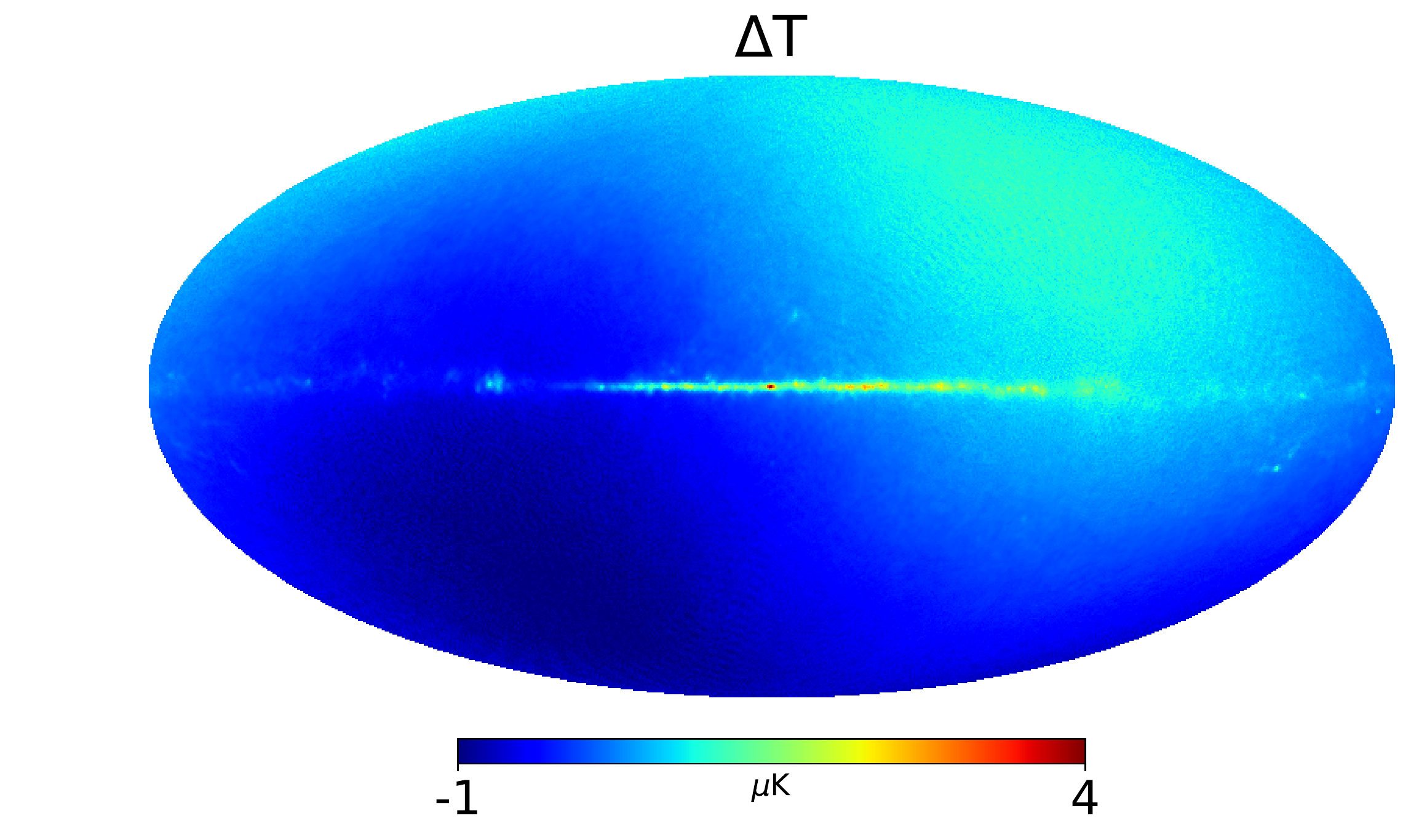}
  \includegraphics[scale=0.28,trim=1.4cm 0 .5cm 0,clip=True]{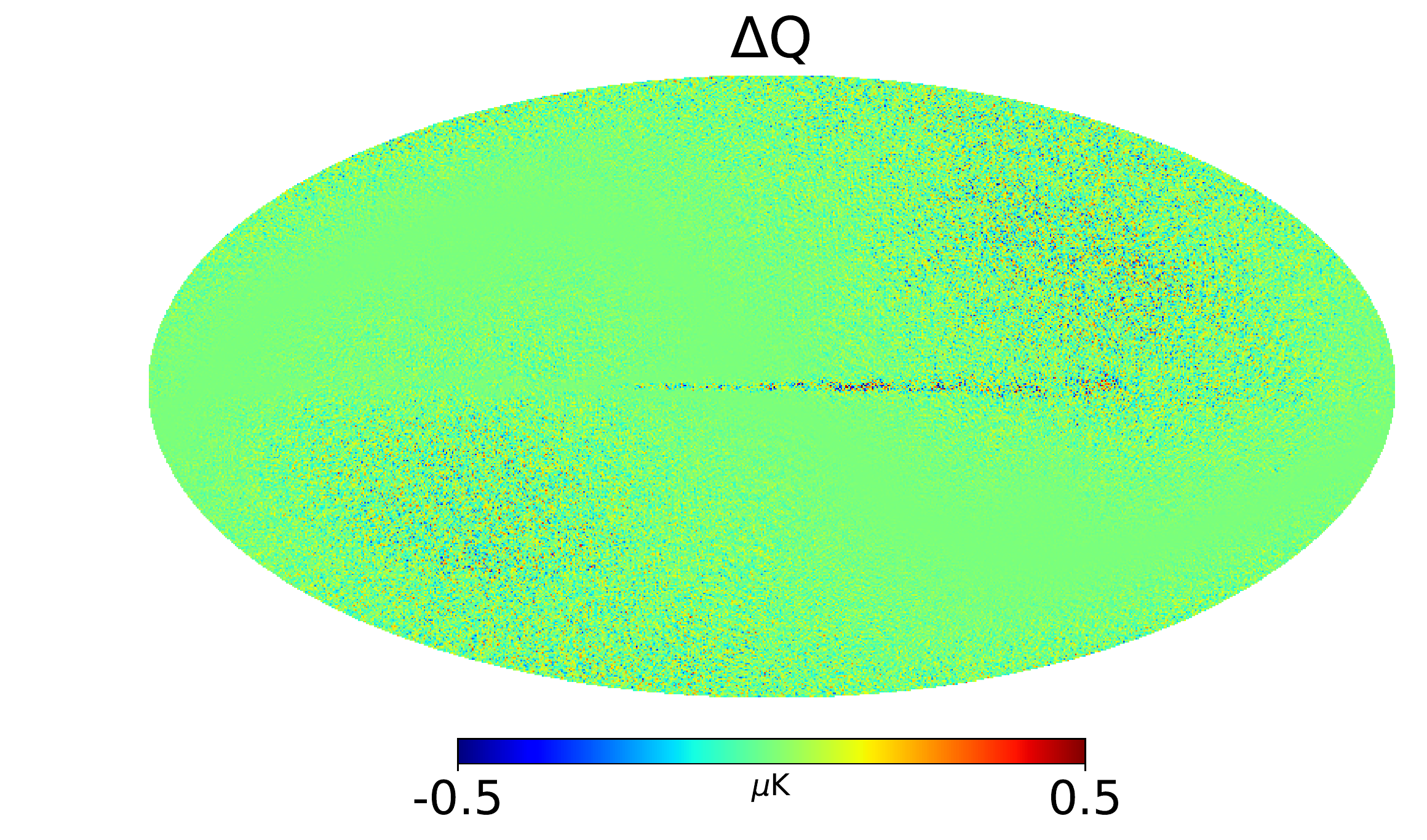} 
  \includegraphics[scale=0.28,trim=1.4cm 0 .5cm 0,clip=True]{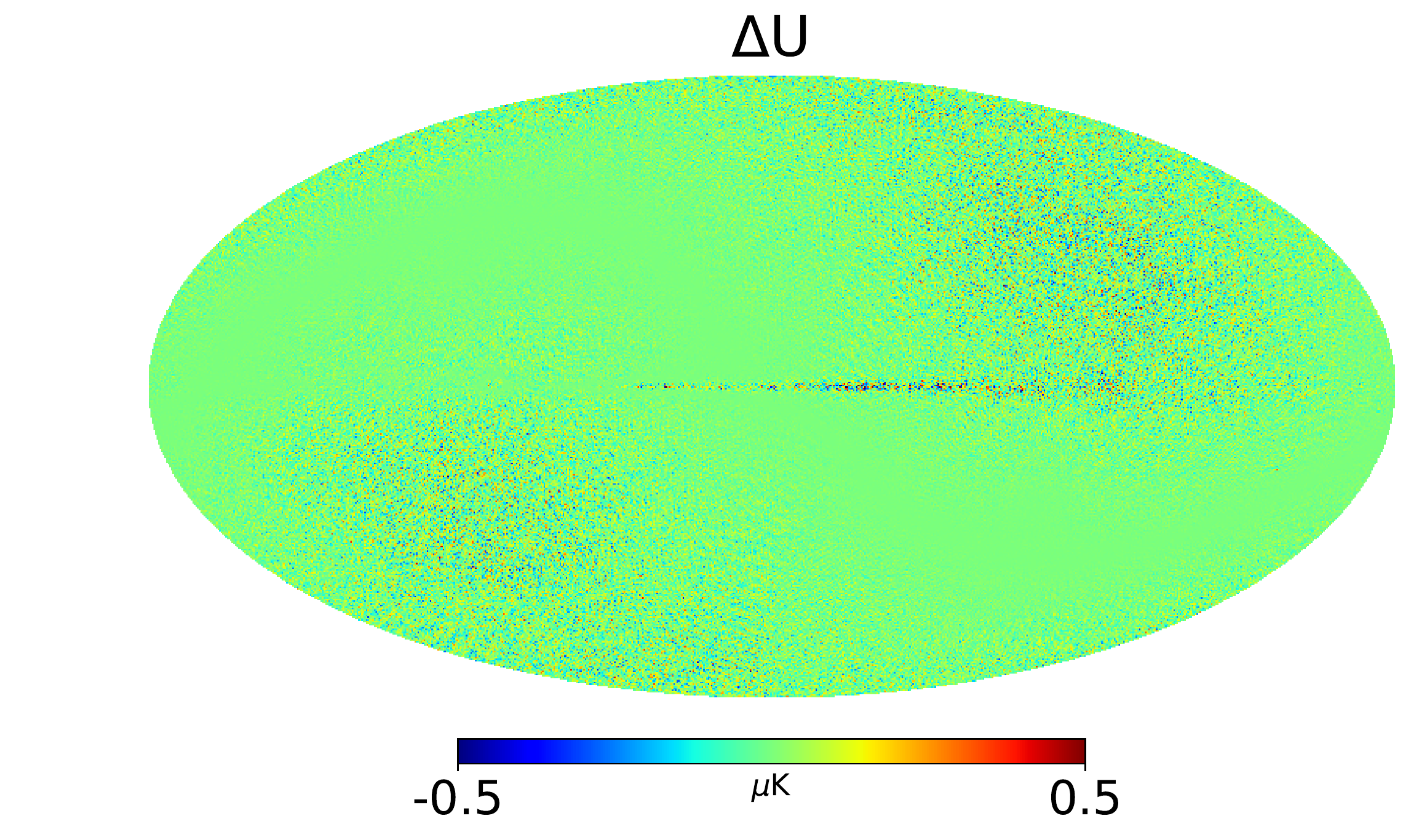}
  \caption{\HW{Top panel: T,Q and U maps reconstructed through a Stokes polarimeter where the HWP wobbles. Bottom panel: Difference between input and output T,Q and U maps showing the effect of the wave-plate precession on observed maps.}}
  \label{fig:diff_maps}
\end{figure}

We perform several simulations with different configurations for the HWP. We vary the spin frequency, precession angle and $I_s/I_{\perp}$ ratio. For each simulation we compare input and output maps and compute the B-modes power spectrum. As visual example, we show in \figurename~\ref{fig:diff_maps} (top panel) the output maps for a mission adopting a \HW{LiteBIRD-like} scanning strategy and solving the Stokes parameters through an HWP with a spin frequency of $1Hz$, a precession angle of $\theta_0=1^\circ$ and $I_{\perp}/I_s=\HW{0.514}$.

The residual maps (i.e. difference between output and input maps) in \figurename~\ref{fig:diff_maps} (bottom panel) show the effect of the HWP wobble that induces variations of few percent with respect to the input map. 
The effect is noticeable close to the galactic plane and close to the maximum and minimum of the solar dipole, where the intensity emission is larger.

\begin{figure}[ht]
\centering
  {\includegraphics[scale=0.28,trim=2.25cm 0.2cm .50cm 0,clip=True]{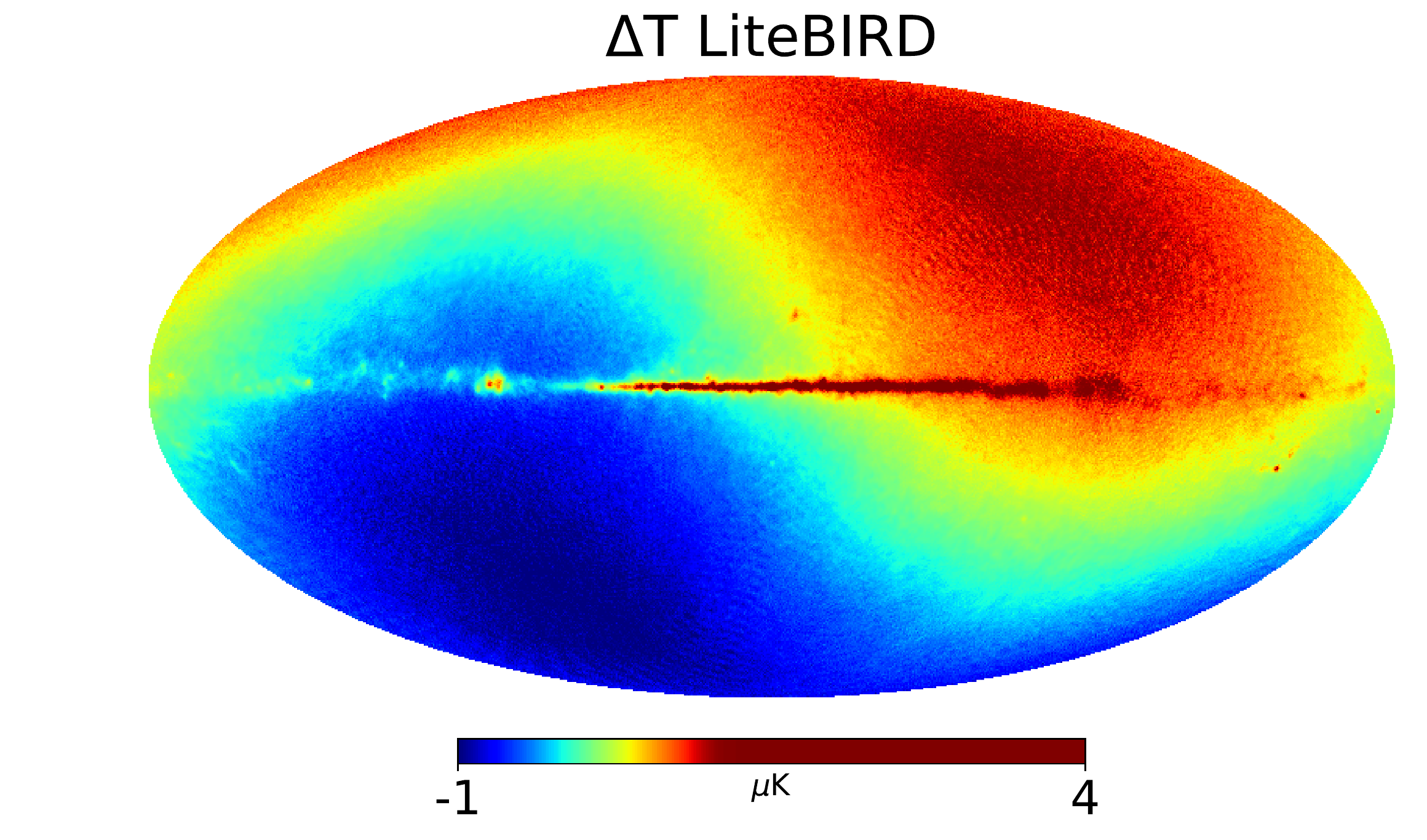}} \quad  
  {\includegraphics[scale=0.28,trim=2.25cm 0.2cm .50cm 0,clip=True]{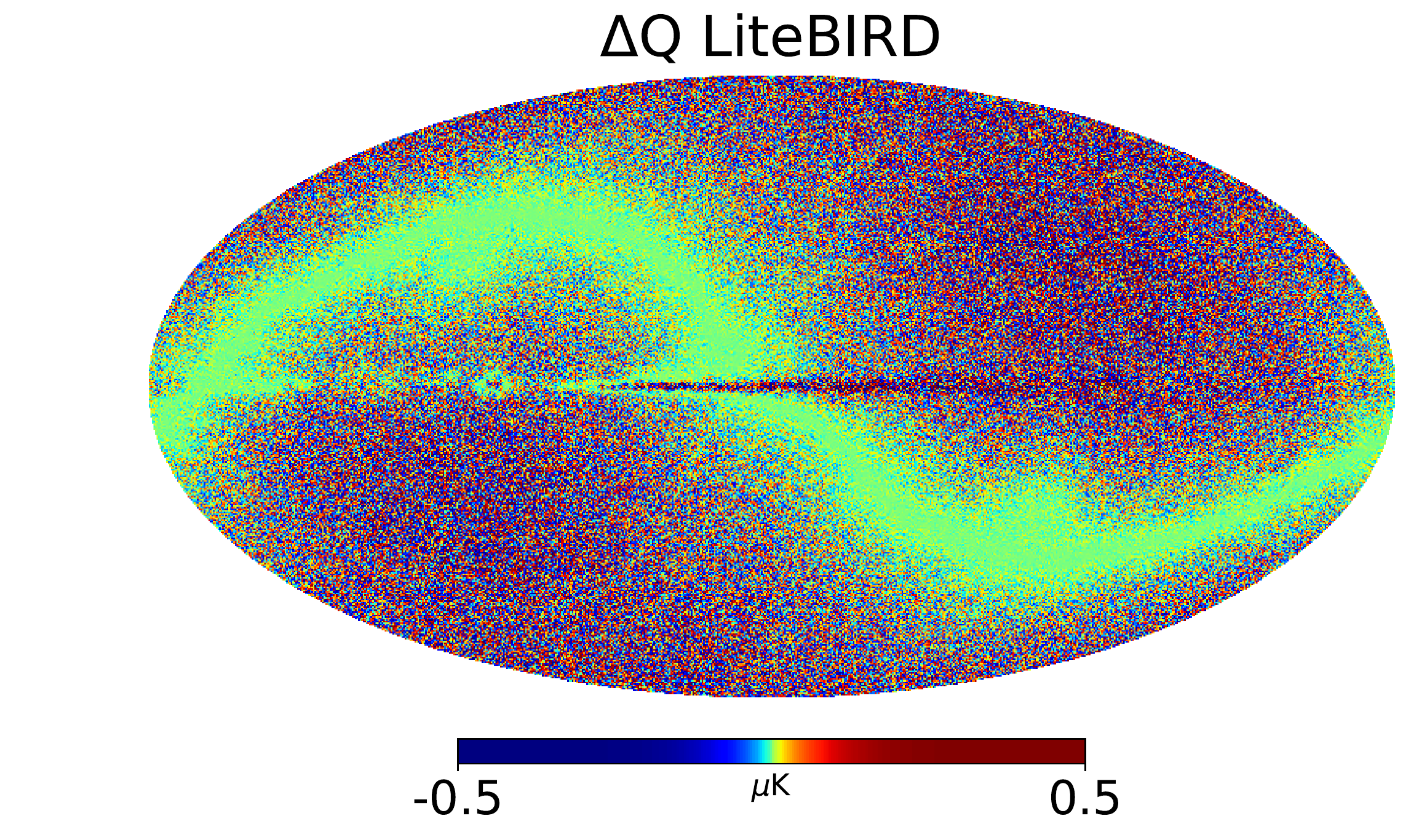}} \quad
  {\includegraphics[scale=0.28,trim=2.25cm 0.2cm .50cm 0,clip=True]{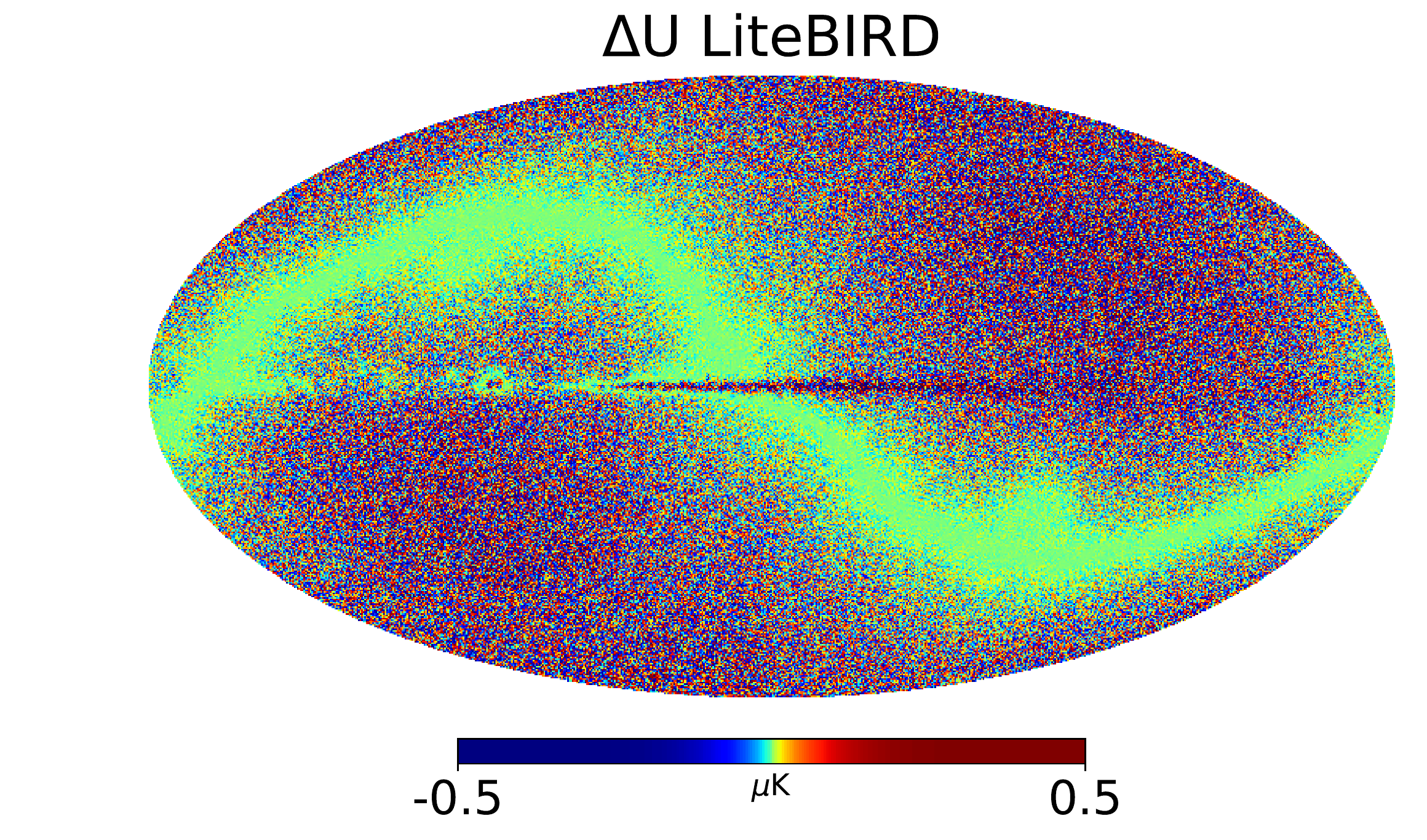}} \quad
  \caption{\HW{Temperature  and polarization difference maps showing the effect of the wave-plate precession for a slow spinning modulator. Modulation parameters:} $f_s=0.1Hz$, $\theta_0 = 1^\circ$ , $I_\perp/I_s=0.514$.} 
  \label{fig:diff_maps_slow}
\end{figure}

Since the effect on the maps is generated by the coupling between the satellite spin and the polarization modulation, affected by the precession, we decided to test several conditions. In particular, slowing the HWP spin speed down to 0.1Hz the effect of the precession is emphasized as you can see in \figurename~\ref{fig:diff_maps_slow}, where the systematic effect induced in the T, Q and U maps, reported in histogram equalized color scale, is at the same level of the input map.

\subsubsection{Results}
\label{sec:results}

\begin{figure}[ht]
\centering
  (a){\includegraphics[scale=0.4]{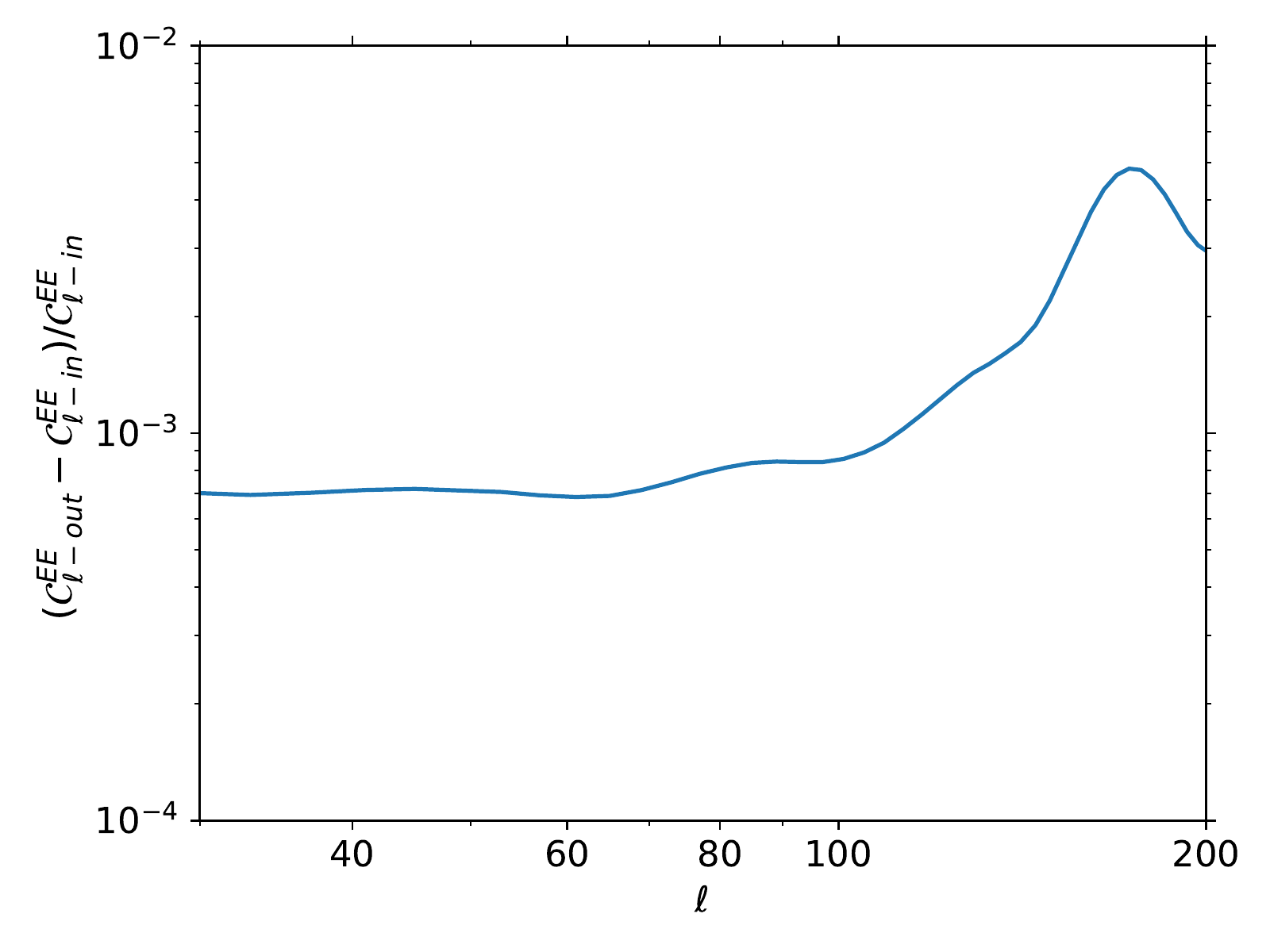}} \quad
  (b){\includegraphics[scale=0.4]{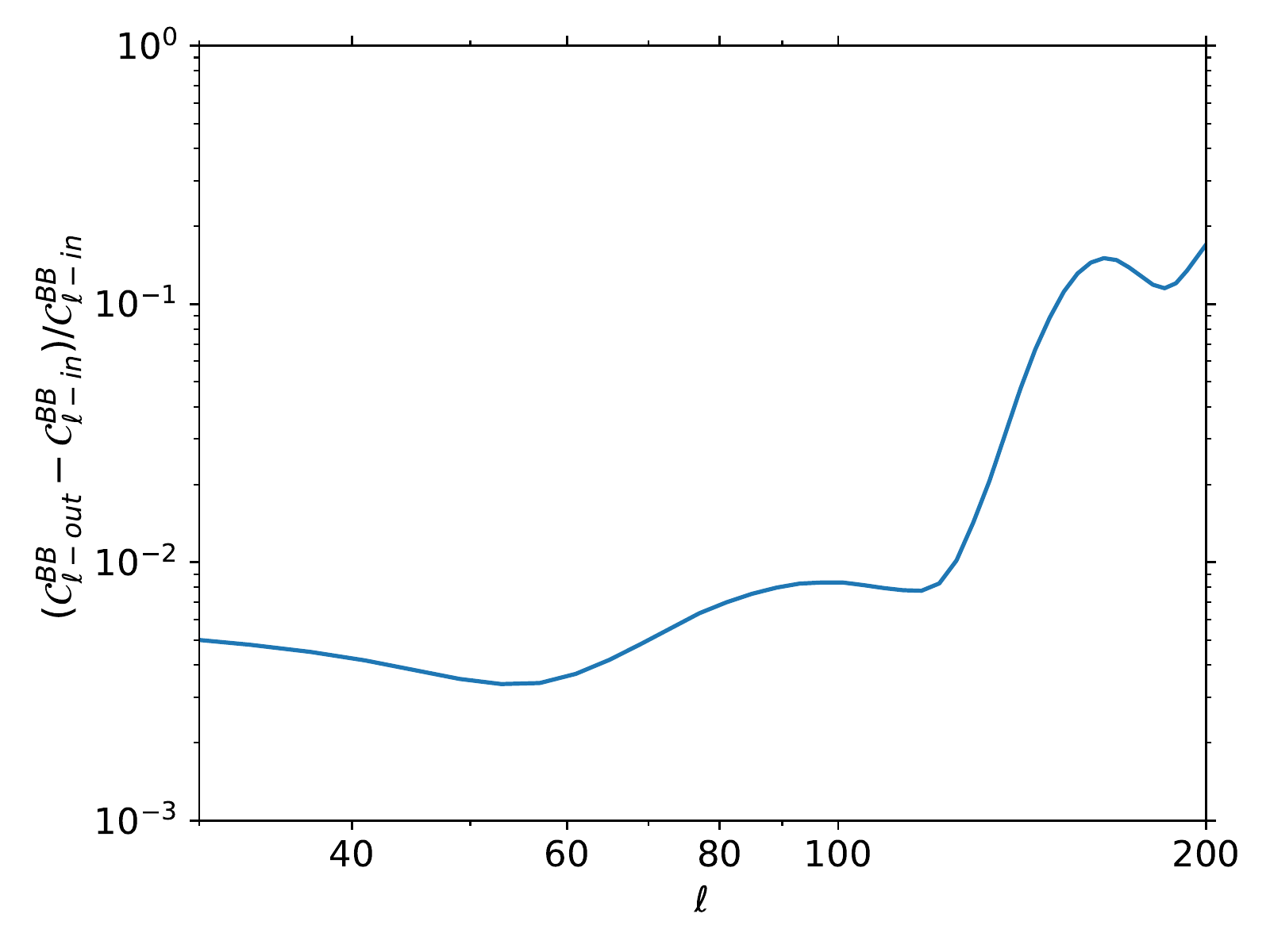}} \quad  
  \caption{Input minus output relative difference of EE and BB power spectra computed from the maps shown in \figurename~\ref{fig:diff_maps} .} 
  \label{fig:power_out}
\end{figure}

The angular power spectra from the output maps shown in \figurename~\ref{fig:diff_maps} are reported up to $l\sim 200$, given the limit imposed by the beam. \HW{The relative variations, for both E and B-modes (\figurename~\ref{fig:power_out}), show the effect of the precession, combined with the satellite spin, that dominates at small angular scales ($l>150$).}

\begin{figure}[ht]
\centering
  (a){\includegraphics[scale=0.4]{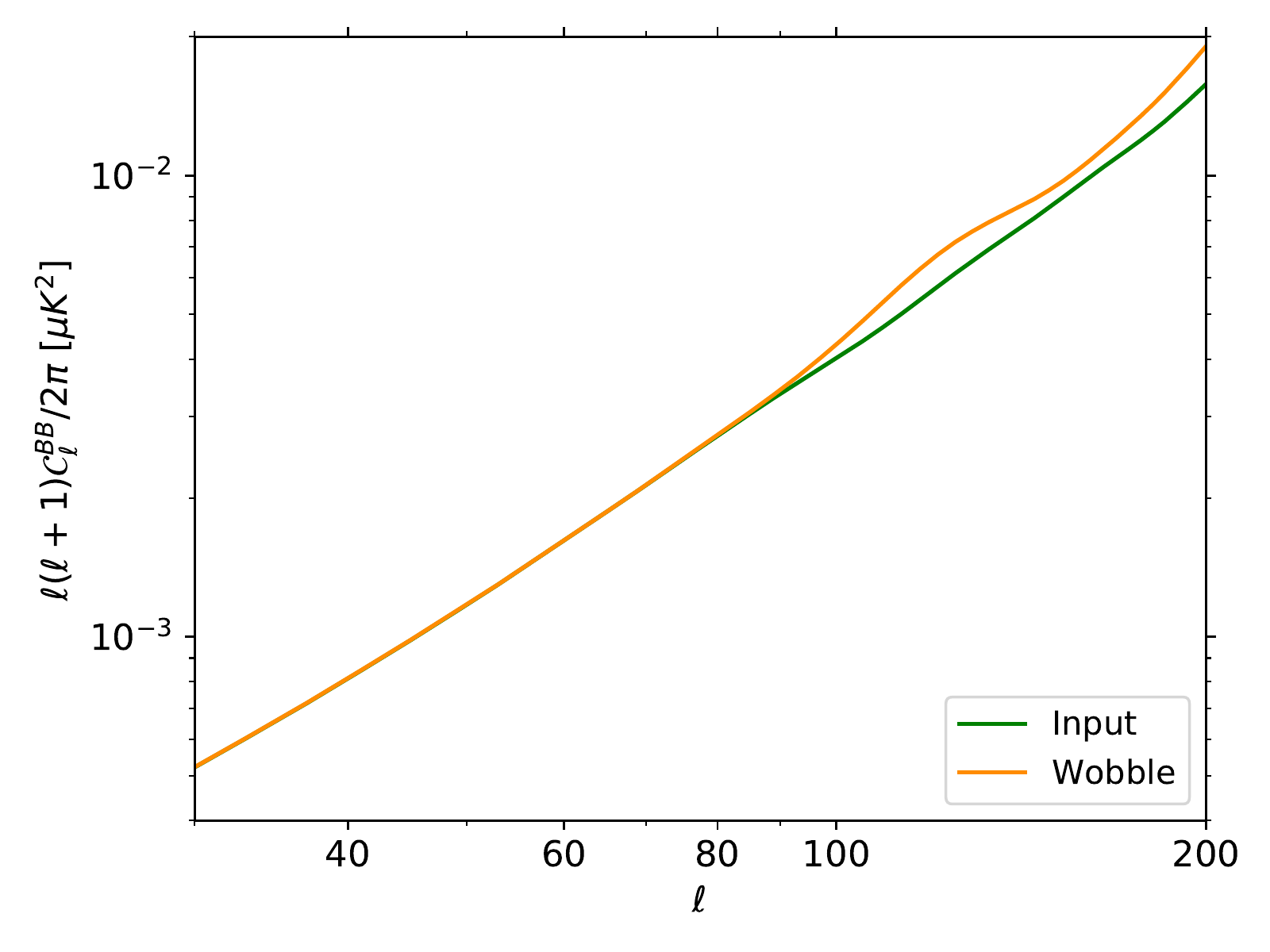}} \quad
  (b){\includegraphics[scale=0.4]{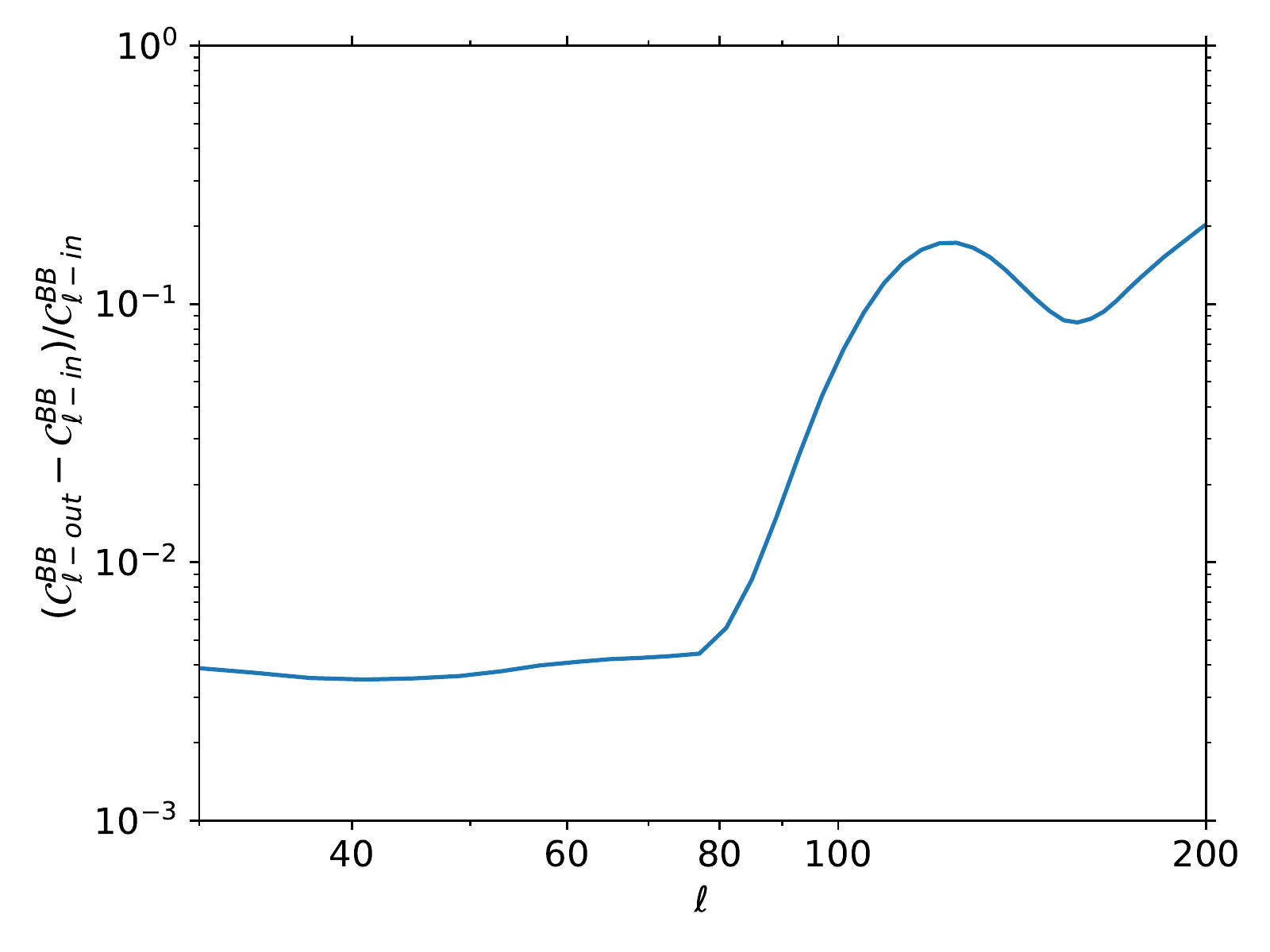}} \quad  (c){\includegraphics[scale=0.4]{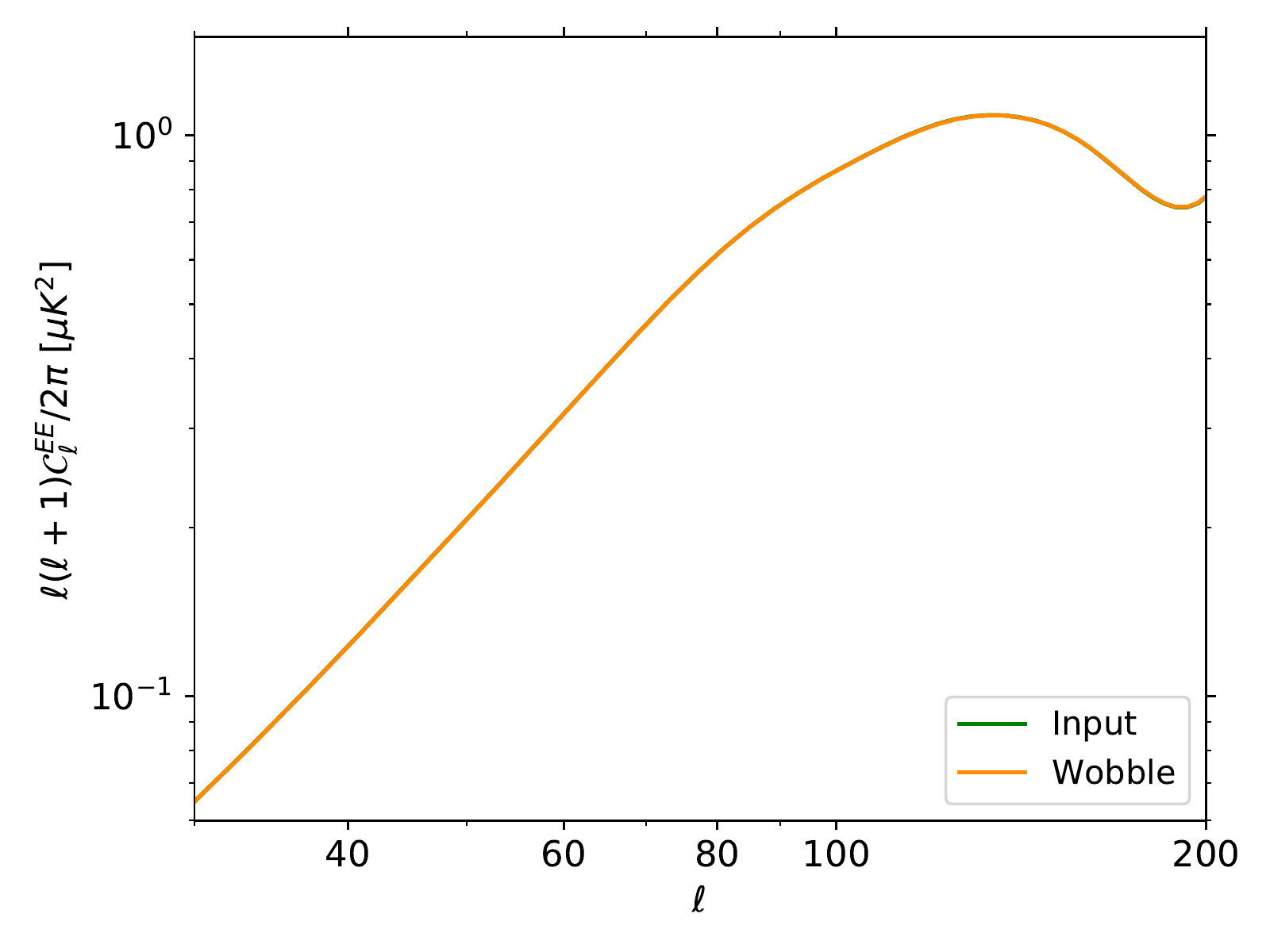}} \quad
  (d){\includegraphics[scale=0.4]{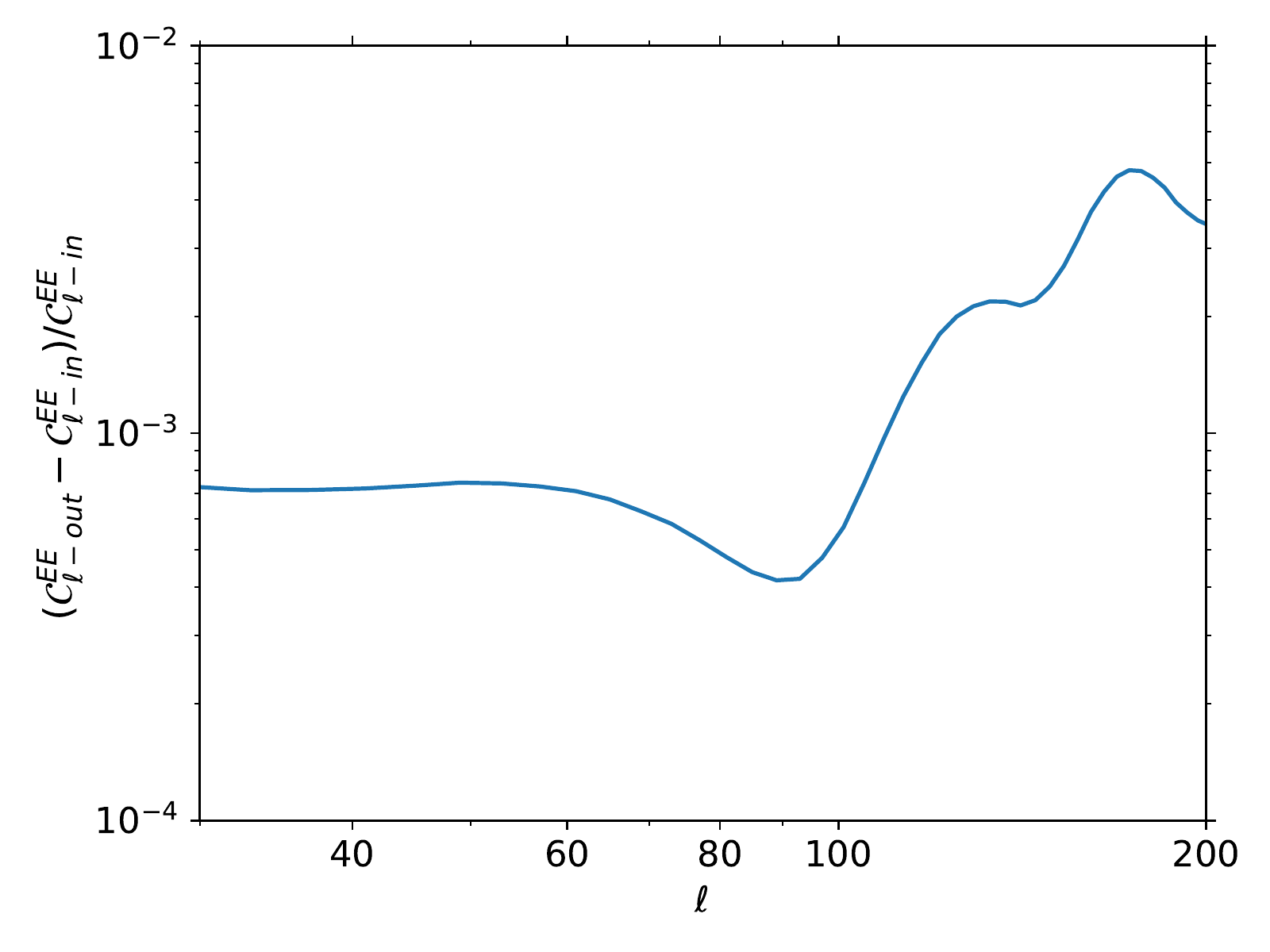}} \quad
  \caption{\HW{(a)-(c) BB and EE Power Spectra extracted from the input map and from the output map of \figurename~\ref{fig:diff_maps_slow}, in (c) the two lines are overlapped. (b)-(d) Normalized difference between input and output power spectra.}} \label{fig:outB_E}
\end{figure}

\HW{If a $0.1$Hz spinning HWP is used, the synchronism with the satellite is slightly different and spurious B-mode polarization shows up at different angular scales ($l>100$). What differs in these two cases is the matching between the systematic effect and the satellite spin (\figurename~\ref{fig:outB_E}-(a)(b)).}

\subsection{Scan strategy comparison}
\label{sec:scan_strategy}
\begin{figure}[ht]
\centering
  \includegraphics[scale=0.20,trim=1.4cm 0 .5cm 0,clip=True]{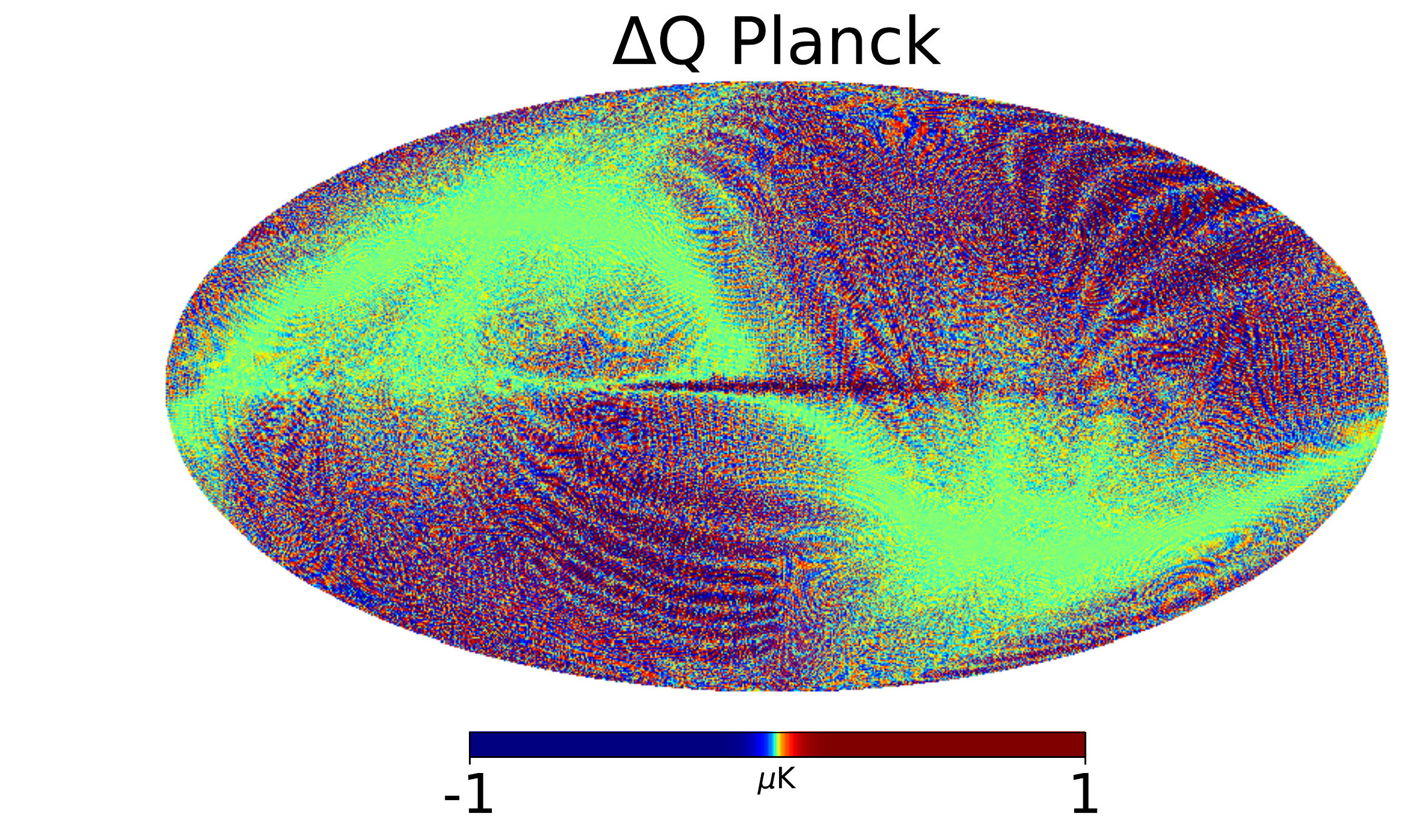}
  \includegraphics[scale=0.20,trim=1.4cm 0 .5cm 0,clip=True]{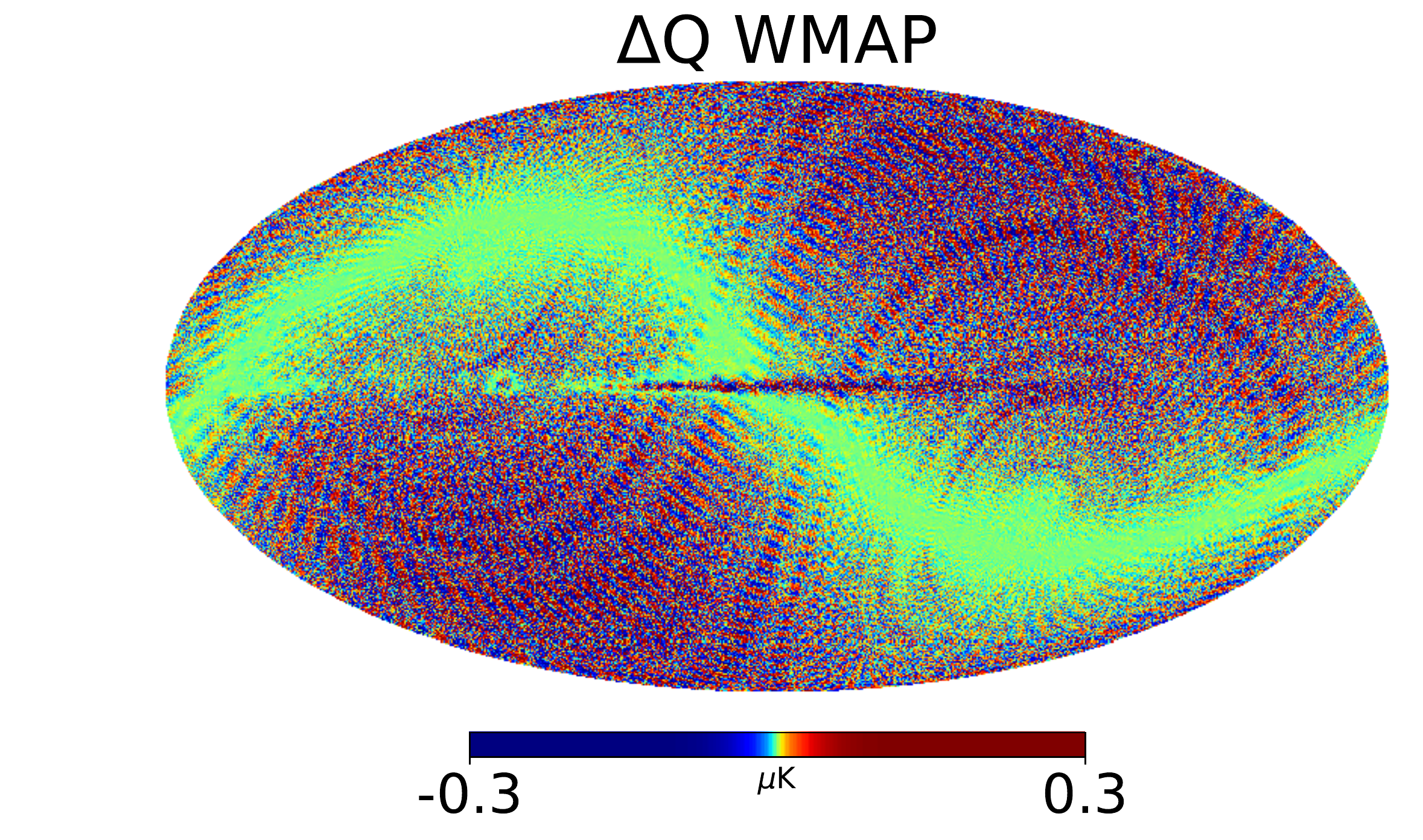} 
  \includegraphics[scale=0.20,trim=1.4cm 0 .5cm 0,clip=True]{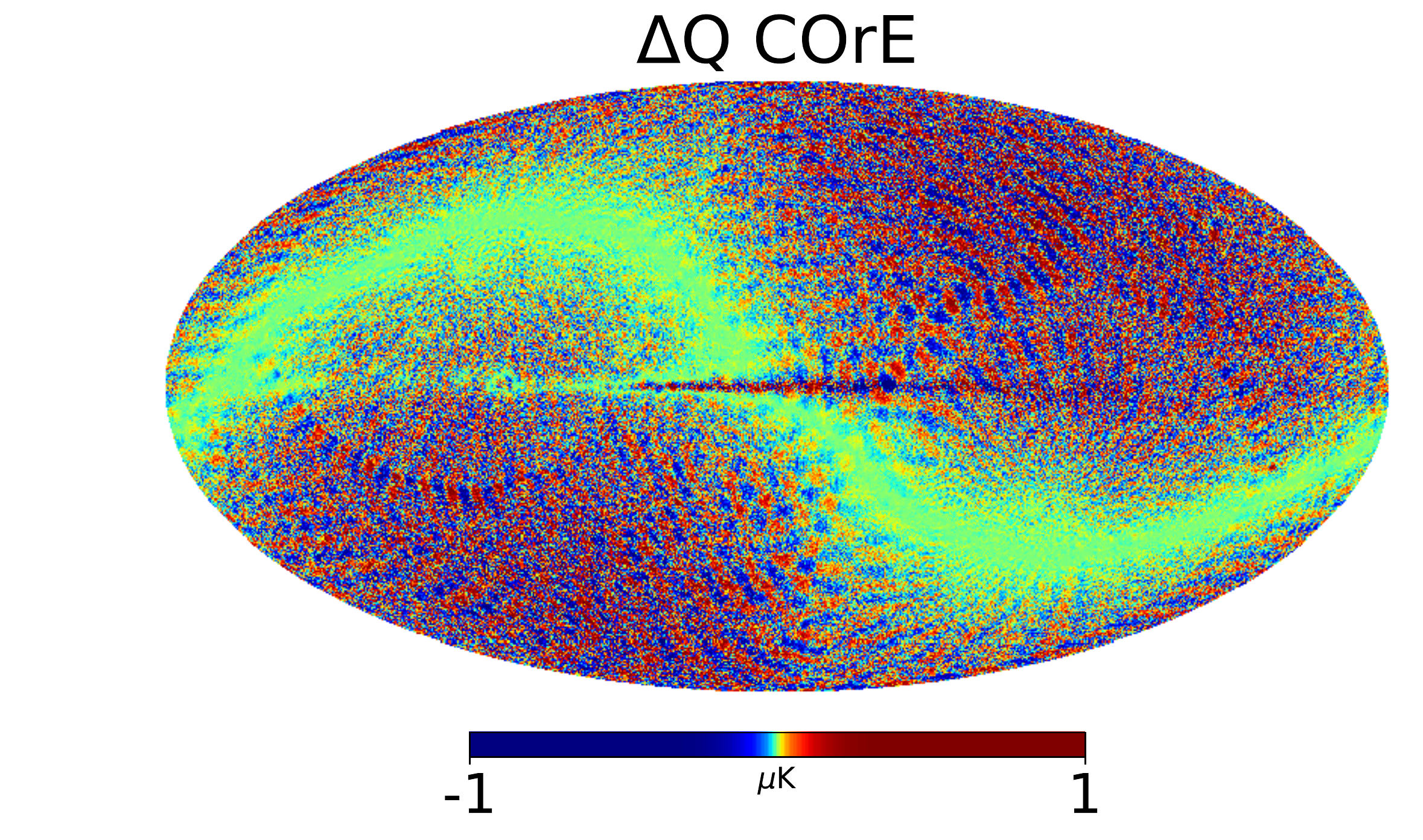}
  \includegraphics[scale=0.20,trim=1.4cm 0 .5cm 0,clip=True]{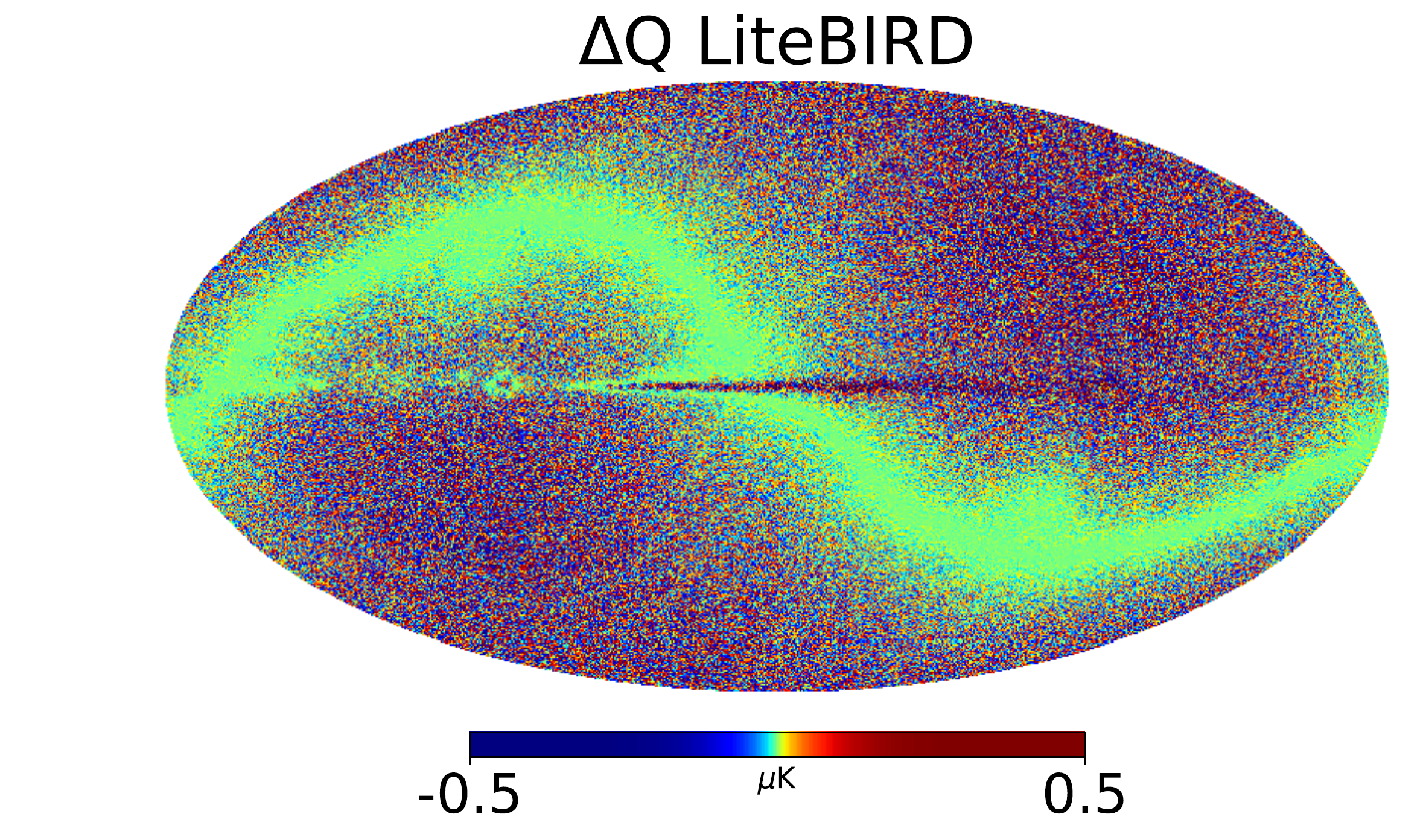}
  \caption{\HW{Difference between polarization Q-maps reconstructed through a Stokes polarimeter showing the direct effect of the wave-plate precession with different scan strategies. Modulation parameters: $f_s=0.1Hz$, $\theta_0=\SI{1}{\degree}$, $I_\perp/I_s=0.508$ for $\planck$, $I_\perp/I_s=0.514$ for \core and LiteBIRD and $I_\perp/I_s=0.510$ for \WMAP . Scan strategy from the left to the right: \planck, \wmap, \core, LiteBIRD.}} 
  \label{fig:out_strategy}
\end{figure}

\begin{figure}[ht]
\centering
(a) {\includegraphics[scale=0.52]{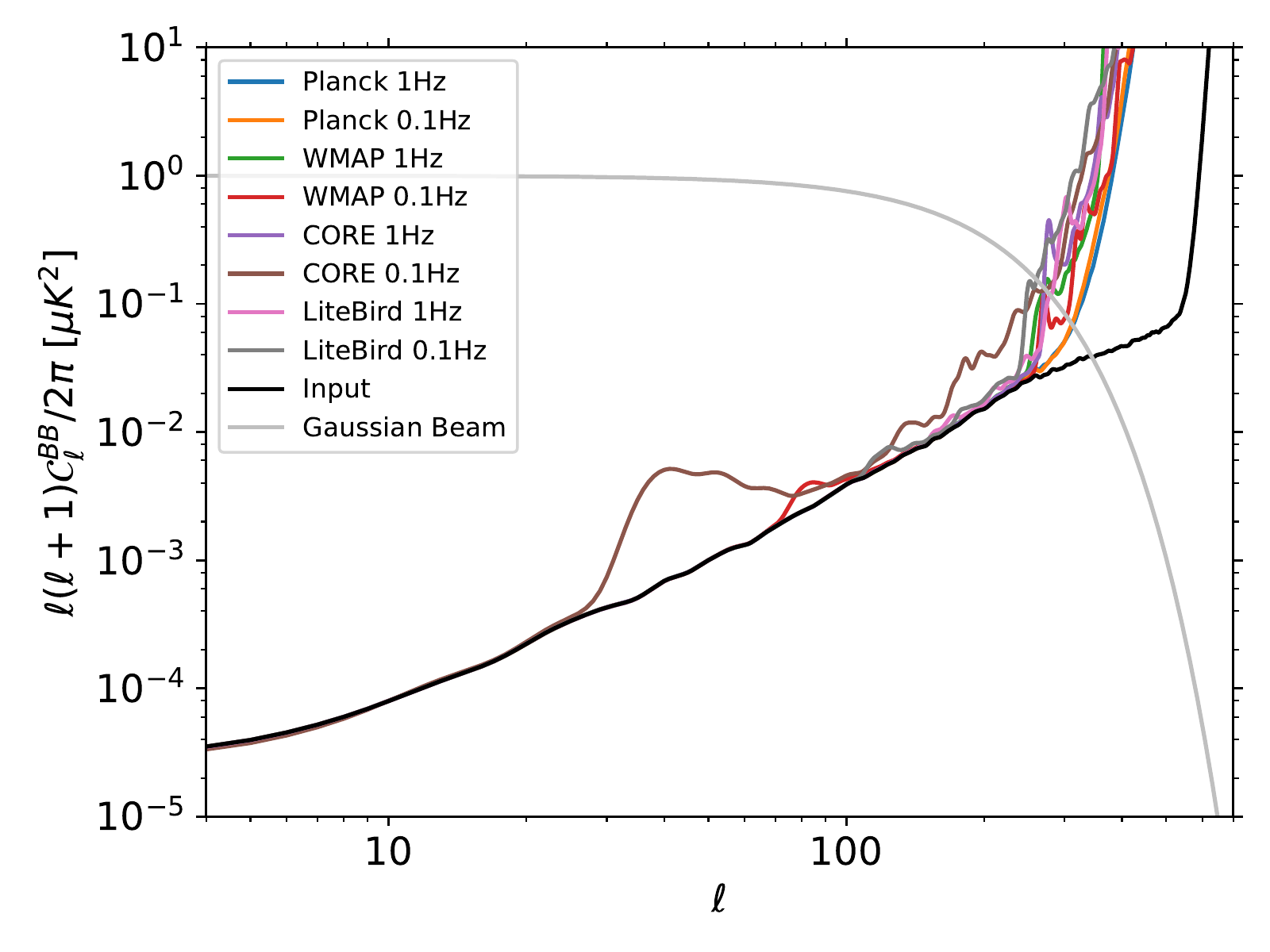}} \quad
(b) {\includegraphics[scale=0.52]{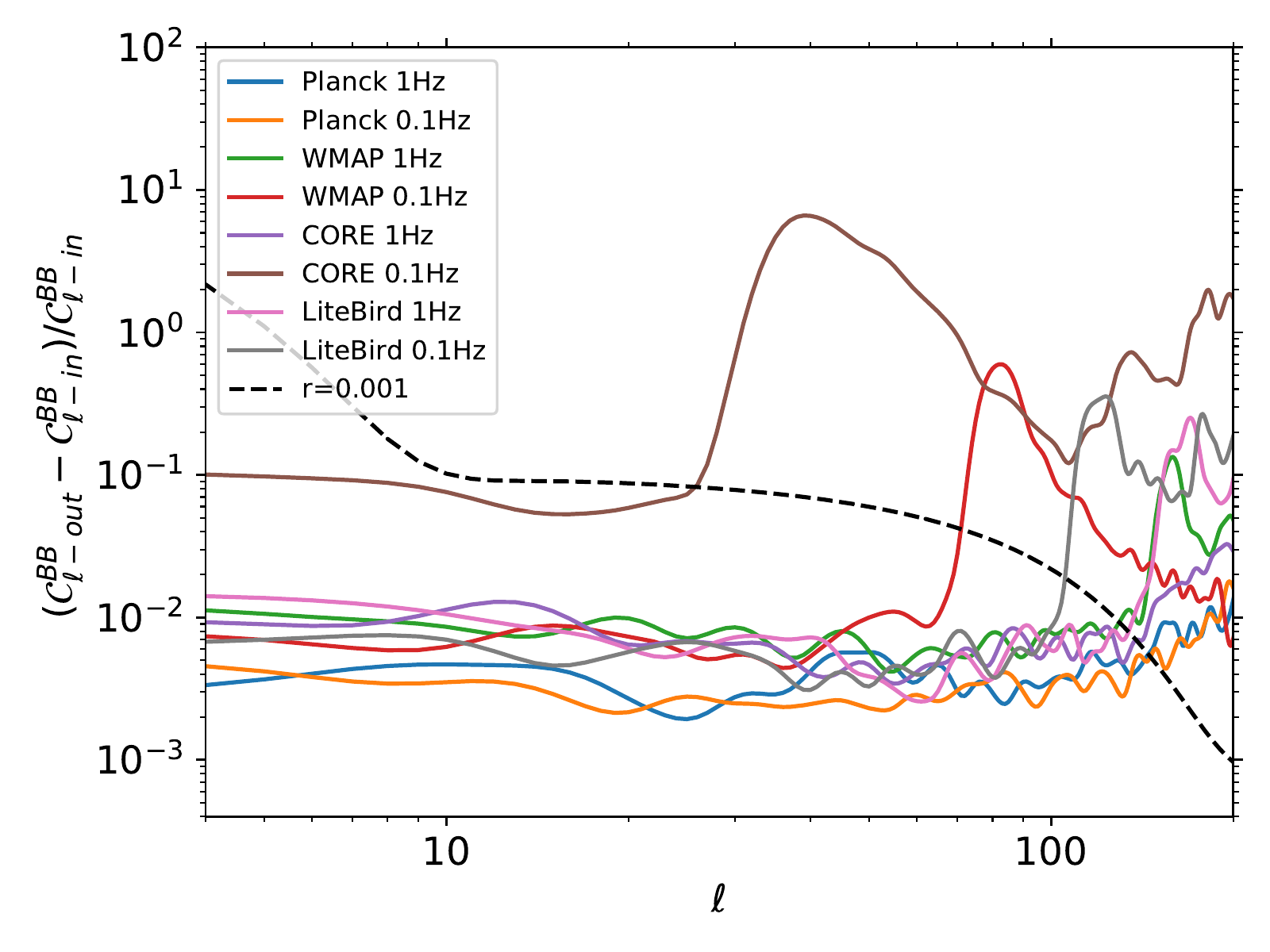}} \quad

  \caption{\HW{B-mode power spectra (a) recovered from simulations including the systematic and the difference output minus input of BB power spectra (b) normalized respect to the input spectrum, when using the \Planck, \WMAP, \core\ and LiteBIRD scan strategies with different HWP spinning speeds ($f_s=[0.1, 1]Hz$). 
  The dashed line represents the primordial B-modes 
  angular power spectrum in the case $r = 0.001$, 
  divided by the lensing only B-modes, assuming the other
  cosmological parameters from \citet{planck2018_overview}.
  }} \label{fig:power_strategy}
\end{figure}

The few simulations presented so far, assuming a \HW{LiteBIRD-like} scanning strategy, show the effect of the HWP precession on full-sky maps and angular power spectra. Since the scan strategy can have a role in mitigating this systematic effect that couples temperature and B-mode polarization \citep{Wallis:2016bja}, we implemented simulations, as described in sec.~\ref{sec:simulator}, able to reproduce different satellite observational strategies.
The results obtained analyzing those simulations are reported as residual maps (i.e. difference between output and input maps), shown in \figurename~\ref{fig:out_strategy}, as root mean square (hereafter RMS) of the residual maps, shown in \tablename~\ref{tab:out_params}, 
and as B-modes angular power spectra, shown in \figurename~\ref{fig:power_strategy}. 

In \figurename~\ref{fig:out_strategy} we report the Q residual maps, in histogram equalized color scale, for the case $f_s=0.1Hz$, $\theta_0=1^\circ$ (the U maps show variations with a similar pattern and similar dynamic range).\HW{ We made the simulations with several HWP physical parameters. Here we report the results for the following values of the $I_{\perp}/I_{s}$ ratio as representative cases: $0.508$ for \planck, $0.514$ for \core  and LiteBIRD and $0.510$ for \WMAP. }    

The result of this analysis shows that the residual does not depend only on the scanning strategy, but mostly on the combination of scanning 
strategy, HWP rotation speed \HW{and $I_\perp/I_s$ ratio}. 

For example, \HW{all} scanning strategy simulations show the largest effect in the case of an HWP spinning at 0.1\,Hz, while they show the lowest residual in the case of 1\,Hz spin frequency. 
This is also evident from the RMS value, reported in \tablename~\ref{tab:out_params}, and from the angular power spectrum in \figurename~\ref{fig:power_strategy}. 
\HW{Anyway, some strategies produce a spurious peak in the angular power spectrum possibly induced by a resonance between satellite spin and HWP wobbling, i.e. \core\ $0.1Hz$ at $\ell\sim40$, \WMAP\ $0.1Hz$ at $\ell\sim85$ or LiteBIRD $0.1Hz$ at $\ell\sim120$.}

\HW{The \Planck-like scanning strategy \citep{Planck:2006aa} does not create particular patterns or structures on larger angular scales, as can be seen in the BB power spectra orange
and blue colored in \figurename~\ref{fig:power_strategy}. }

On the other hand the \HW{\core-like} simulation (with slowly spinning HWP) shows the worst coupling between the satellite spin and wave-plate precession in terms of angular structures at large scales, as visible in terms of spurious B-modes (\HW{brown} colored in \figurename~\ref{fig:power_strategy}).
These simple cases show that the large scale patterns arisen in the residual map are not related to the whole quality of the map better \HW{described by the angular power spectrum}.

The power spectra and the maps recovered show the \HW{contaminations} generated by the half-wave plate precession systematic, for a specified scan strategy. 
Repeating the analysis with different precession angles we 
conclude, as expected, that the larger is the precession angle, the larger is the spurious B-mode signal; the higher is the HWP spin frequency, the greater is the mitigation of the systematic effect. 

In terms of research for primordial B-modes, the faster rotation of the wave-plate \HW{helps} to mitigate the systematic \HW{effect} induced by the precession of the modulating element in a Stokes polarimeter \HW{by moving the contamination at high $\ell$}.   
Fig.\ref{fig:power_strategy}-(b) illustrates  the fractional residual B-modes due to observation  with a wobbling HWP, in the case of no-tensor  perturbations, but only lensing-induced B-modes. The fractional residual power is a good figure of merit of the contamination, given that next-generation CMB polarization experiment are designed to reach a sensitivity which is usually quantified as a fraction of the lensing-induced B-modes level.

\begin{table}
\centering
\begin{tabular}{c || c | c | c | c | c | c | c | c | c | c | c | c } 
 {} & \multicolumn{3}{c|}{\Planck} & \multicolumn{3}{c|}{\WMAP} & \multicolumn{3}{c}{\core} & \multicolumn{3}{|c}{LiteBIRD} \\
 \hline
 {Frequency [Hz]} & {T[\SI{}{\micro\kelvin}]} & {Q[\SI{}{\micro\kelvin}]} & {U[\SI{}{\micro\kelvin}]} & {T[\SI{}{\micro\kelvin}]} & {Q[\SI{}{\micro\kelvin}]} & {U[\SI{}{\micro\kelvin}]} & {T[\SI{}{\micro\kelvin}]} & {Q[\SI{}{\micro\kelvin}]} & {U[\SI{}{\micro\kelvin}]} &
 {T[\SI{}{\micro\kelvin}]} & {Q[\SI{}{\micro\kelvin}]} & {U[\SI{}{\micro\kelvin}]}\\
 \hline
 {1.0}	&	{0.600}	&	{0.032}	&	{0.032}	&	{0.601}	&	{0.071}	&	{0.071}	&	{0.600}	&	{0.071}	&	{0.071} &	{0.600}	&	{0.071}	&	{0.071}\\
 \hline
 {0.1}	&	{0.600}	&	{0.050}	&	{0.050}	&	{0.600}	&	{0.072}	&	{0.072}	&	{0.600}	&	{0.088}	&	{0.087} &	{0.600}	&	{0.076}	&	{0.076} \\
\end{tabular}
\caption{\HW{RMS values in \SI{}{\micro\kelvin} for intensity and polarization for different scan strategies. Maps correspond on the input/output difference maps for each case.}} 
\label{tab:out_params}
\end{table}

\subsection{Temperature to Polarization leakage}

Including in the input for the simulations a map with only temperature foregrounds, we can highlight the temperature-to-polarization leakage effect induced by the systematic for various scenarios. We verified that polarization foregrounds, removed with ideal component separation method, leave one order of magnitude lower residual in terms of P-P leakage.

\begin{figure}[ht]
\centering
{\includegraphics[scale=0.50]{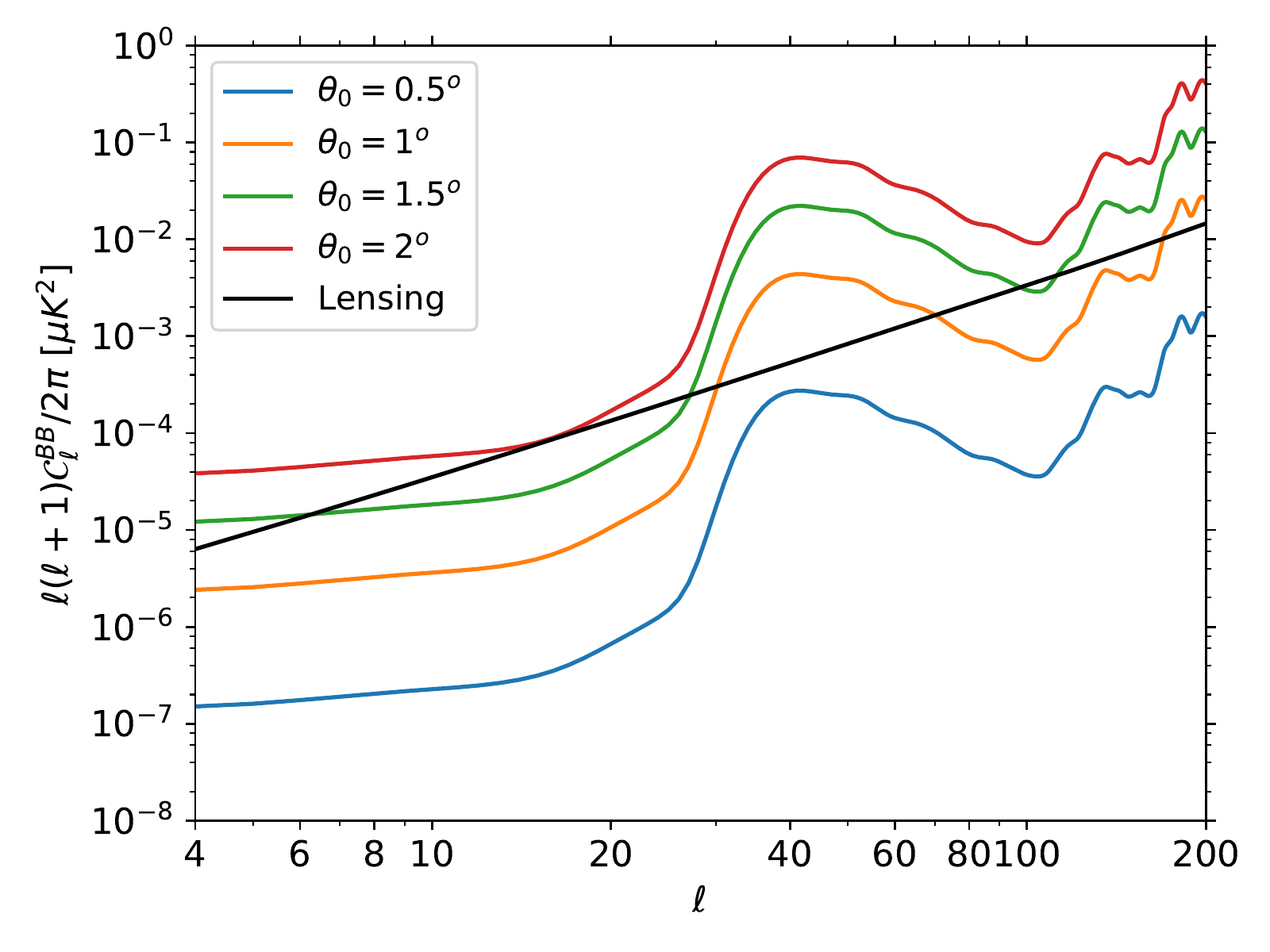}} \quad

  \caption{\HW{B-mode power spectra recovered from simulations including the systematic when using the \core\ scan strategy with fixed HWP spinning speed ($f_s=0.1Hz$) and different precession angles. The input map includes only temperature contributions, the polarization contribution in the output map arises from the HWP precession.}} 
  \label{fig:power_noPol}
\end{figure}

The HWP wobble induces B-modes which amplitude is proportional to $\theta_0$ as you can see in \figurename~\ref{fig:power_noPol} where we report the BB power spectra for different precession angles [$0.5^\circ$, $1^\circ$, $1.5^\circ$, $2^\circ$] extracted from the maps scanned by a \HW{\core-like} satellite. The induced B-mode signal exceeds the gravitational lensing contribution already for $\theta_0=1^\circ$.

The output polarization components Q and U are shown in Table~\ref{tab:out_params_strategy}.  

\begin{table}
\centering
\begin{tabular}{c | c | c | c} 
 $\theta_0$ [$^\circ$] & T[\SI{}{\micro\kelvin}] & Q[\SI{}{\micro\kelvin}] & U[\SI{}{\micro\kelvin}] \\
 \hline
  $0.5$  & 0.150 & 0.533 & 0.531 \\
  $1$    & 0.600 & 0.540 & 0.537 \\
  $1.5$  & 1.350 & 0.569 & 0.569 \\
  $2$    & 2.401 & 0.640 & 0.640 \\
\end{tabular}
\caption{\HW{RMS values in \SI{}{\micro\kelvin} for intensity and polarization residual maps considering a \core-like scan strategy with the HWP spin speed at 0.1Hz and with different precession angles.}} 
\label{tab:out_params_strategy}
\end{table}

\subsection{Neighbouring Detector To Mitigate The Systematic}
\label{sec:neighboring_pixels}
  The systematic effect induced by the wobbling can be mitigated by redundancy. Observing the same sky pixel with different phases of the wobbling 
  plate averages out the contamination. 
  This can be also obtained by the combination of multiple detectors, 
  observing the same pixel at different times. In order to check 
  this mitigation, we simulate the observation with two different 
  detectors, pointing to different boresight angles $\beta$, shifted by 
  1$^\circ$, for all the proposed scanning strategies. 
  The modulation parameters used are:$f_s=1Hz$, $\theta_0=1^\circ$, \HW{$I_\perp/I_s=0.508$ for \planck, $I_\perp/I_s=0.514$ for \core and LiteBIRD and $I_\perp/I_s=0.510$ for \WMAP.}
  The combination of the data from the two detectors
  results in a single map with a reduced contamination
  with respect to the single detector maps, as reported in Table 
  \ref{tab:pixel_strategy}. In the Table we report 
  the RMS of the difference between the map with and without the
  induced systematic effect. This RMS of the residual is 
  very similar for the two single-detector maps, and is reduced 
  in the map produced with their combination. The RMS of the residuals scales 
  with a factor $\sim \sqrt 2$, indicating that the contamination is rather uncorrelated among the two detectors.

\begin{table}
\centering

\begin{tabular}{c | c | c | c} 
 & $\beta$ [$^\circ$]  & Q[\SI{}{\nano\kelvin}] & U[\SI{}{\nano\kelvin}] \\
 \hline
 \hline
 \Planck & $85.0$     & 31.9 & 32.0 \\
 		& $86.0$     & 31.8 & 31.8 \\
\hline
        & combination     & 22.6 & 22.6 \\
\hline
\hline
 \WMAP\   & $70.5$   & 70.7 &  70.8 \\
 		& $71.5$     & 71.8 & 71.8 \\
 \hline
        & combination    &  49.8 &  49.9 \\
\hline
\hline
 \core\	& $65.0$   & 71.0 &  70.7 \\
		& $66.0$     & 72.6 &  72.6 \\
\hline
        & combination     & 50.9 & 50.5 \\
\hline
\hline
 LiteBIRD\	& $50.0$   & 70.9 & 70.9  \\
	     	& $51.0$   & 70.1 &  70.1\\
\hline
        & combination     & 50.1 & 50.1 \\
\end{tabular}
  \caption{\HW{RMS values, in \SI{}{\micro\kelvin}, of the difference 
  between the maps with and without the effect of the HWP wobbling. 
  The RMS is calculated for the maps from the two detectors, and 
  for the map from the two detectors combined. 
  The RMS of the residual is reduced by a factor $\sim \sqrt{2}$ 
  combining two detectors. 
  Modulation parameters: $f_s=1Hz$, $\theta_0=1^\circ$, $I_\perp/I_s=0.508$ for \planck\ and \core, $I_\perp/I_s=0.514$ for LiteBIRD and $I_\perp/I_s=0.510$ for \WMAP . } }
  \label{tab:pixel_strategy}
\end{table}

\section{Conclusions}
\label{sec:conclusions}
We described the systematic error induced by the precession of the half-wave plate modulator in a Stokes polarimeter and the effects on CMB polarization measurements.

In section \ref{sec:stokes_polarimeter} we derived the analytical equation (\HW{Eq.~\ref{eq:intensity_woobled_Muller})} for the power on the central detector of the polarimeter when the HWP precesses with an angle $\theta_0$ and with a frequency $\gamma$ imposed by the physical properties of the device (the spin frequency $f_s$ and the $I_\perp/I_s$ ratio, including its support). By using \HW{Eq.~\ref{eq:intensity_woobled_Muller}} we performed several simulations to assess the level of the systematic effect induced by the precession. We found a strong dependence on $\theta_0$ and $\gamma$, both for the fractional residual of the signal and for the power spectrum.

In section~\ref{sec:simulations} we developed the simulation of full-sky observation by a satellite mission, including the systematic effect, and quantified its impact on B-modes retrieved from the output map. The HWP precession produces spurious effects on the maps at different angular scales depending on the strategy used to scan the sky; in particular, we implemented four scan strategies, \WMAP, \Planck, \core\ and LiteBIRD like. 
With the specific mechanical properties of the implemented HWP, 
our simulations show a B-mode induced by leakage from intensity signal that dominates at different scales for the strategies selected: 
\HW{at $\ell\sim 40$ for \HW{\core-like} satellite, at \HW{$\ell \simeq 85$} for a \WMAP-like and \HW{$\ell \simeq 120$} in the LiteBIRD-like case. }
The new-era CMB experiment will gain some order of magnitude in sensitivity \citep{stage4}, few $10^{-4}\mu K$, compared nowadays. By having the analytical formula of the systematic effect induced 
\HW{by the HWP wobbling, it is possible to remove it or, at least, to forecast its impact on data. }

In general, the effect of the precession is to induce beats into the timelines. Those beats, in amplitude and frequency, are related to the physical properties of the moving parts. Their effect into the maps depends on specific scanning parameters, and on possible coupling with the beats frequency.  
We recommend to take this effect into account in the design of an observation strategy, by modelling and measuring the inertia tensor of the moving parts. Once the tensor is measured (or modelled) it can be inserted into \HW{Eq.~\ref{eq:intensity_woobled_Muller}} to simulate the impact into the timelines. The scanning strategy must avoid any synchrony with the beats frequency. In this case, redundancy helps cancelling most of the contamination, but, considering the targeted sensitivity of future CMB experiments \citep{LiteBIRD:18}, the precession must still be considered as a possible source of systematic effect. 

\begin{acknowledgements}
\label{sec:acknowledgements}
We acknowledge the support of the ASI-COSMOS programme, Prof. Elia Battistelli for CPU time on SPINDUST server. The work has been supported by University of Rome ''Sapienza`` research funds. LP acknowledges the support of the CNES postdoctoral program. 
\end{acknowledgements}

\appendix
\section{Variable definitions}

We list in this short appendix the main variables defined in the paper:\\

\noindent $E_k$ = Electric field components k=(x,y,z);\\
$T, Q, U$ = 2D Stokes Parameters;\\
$\Delta_i$ = Component of 3D Stokes vector;\\
$s$ = 3D generic Stokes vector;\\
$s_{in}$ = 3D input Stokes vector;\\
$s_{out}$ = 3D output Stokes vector;\\
$i_{in}$ = Input Jones vector;\\
$i_{out}$ = Output Jones vector;\\
$I$ = Intensity at detector;\\
$\sigma_i$ = Gell-Mann matrix;\\
$m$ = HWP support mass;\\
$h$ = HWP support thickness;\\
$R$ = HWP support radius;\\
$I_s$ = Moment of inertia component along z-axis;\\
$I_\perp$ = Moment of inertia component Along x(y)-axis;\\
$L_k$ = Angular momentum components k=(x,y,z);\\
$f_s = \omega_s/2\pi$ = HWP spin frequency respect to z-axis;\\
$\gamma$ = HWP precession frequency;\\
$\theta$ = HWP orientation angle respect to z-axis;\\
$\theta_0$ = HWP precession angle;\\
$\eta$ = HWP rotation angle respect to x-axis composing the precession;\\
$\xi$ = HWP rotation angle respect to y-axis composing the precession;\\
$r$ = Distance from the optical axis on the detector plane;\\
$\omega_1$ = Earth revolution velocity;\\
$\omega_2$ = Precession velocity in the satellite scanning strategy;\\
$\omega_3$ = Satellite spin;\\
$\alpha$ = Satellite precession angle, i.e. the angle between the satellite spin axis and the sun-earth direction;\\
$\beta$ = Boresight angle, i.e. the angle between the focal plane direction and the spin axis.\\

\section{Precession Theory}
\label{app:pre_teo}
We report in this appendix the derivation of the main equations describing a precessing disc, in particular we detail the derivation of Eq.~\ref{eq:gamma}.\\
The contribution to $L_x$ due to the rotation about the x-axis is $L_x=\frac{d(I_{xx} \eta)}{dt}=I_{xx} \frac{d \eta}{dt} $. We can treat $I_{xx}$ as a constant since moments of inertia about the principal axes are constant for small angular displacements. In addition, the rotation about the $y$-direction contributes to $L_x$ by giving a component $L_s \sin(\xi)$ on the x-axis. Combining such contributions we get:

\begin{eqnarray}
L_x &=& I_{xx} \frac{d \eta}{dt} + L_s \sin(\xi)\;,\nonumber\\
L_y &=& I_{yy} \frac{d \xi}{dt} - L_s \sin(\eta). 
\label{eq:angular_momentum}
\end{eqnarray}

\noindent Since $I_{xx}=I_{yy}=I_\perp$ and exploring small angle first order approximation 
Equations~\ref{eq:angular_momentum} read:

\begin{eqnarray}
	L_x &=& I_\perp \frac{d \eta}{dt} + L_s\xi\;,\nonumber\\
	L_y &=& I_\perp \frac{d \xi}{dt} - L_s\eta.
    \label{eq:angular_momentum2}
\end{eqnarray}

\noindent Furthermore, thanks to the same approximation $L_z=I_s \omega_s$.
Since we are considering a torque-free system ($dL/dt=0$), both $L_s$ and $\omega_s$ are constant leading to:

\begin{eqnarray}
	 I_\perp \frac{d^2 \eta}{dt^2} + L_s\frac{d \xi}{dt}&=&0 \;, \nonumber \\
	 I_\perp \frac{d^2 \xi}{dt^2} - L_s\frac{d \eta}{dt}&=&0.
	\label{eq:motion_equation}
\end{eqnarray}

\noindent By introducing $\omega_x=\frac{d \eta}{dt}$ and $\omega_y=\frac{d \xi}{dt}$ Eq.~\ref{eq:motion_equation} become:

\begin{eqnarray}
I_\perp \frac{d \omega_x}{dt} + L_s\omega_y &=&0\;,\nonumber\\
I_\perp \frac{d \omega_y}{dt} - L_s\omega_x &=&0 
	\label{eq:motion_equation2}
\end{eqnarray}

\noindent In order to solve this coupled system of differential equations we can differentiate one and substitute the other:

\begin{equation}
\frac{d^2 \omega_x}{dt^2} + \gamma^2 \omega_x =0\;\;\;\; {\rm with }\;\;\;\; \gamma=\frac{L_s}{I_\perp}=\omega_s \frac{I_s}{I_\perp}\;.
\label{eq:freq}
\end{equation}

\noindent The solution for the harmonic motion is (with $A$ and $\phi$ arbitrary constants):
\begin{equation}
\omega_x = A \sin(\gamma t + \phi ) \;.
\label{eq:omega_x}
\end{equation}

\noindent While for $\omega_y$ we get:

\begin{equation}
\omega_y =  - \frac{I_\perp}{L_s} \frac{d \omega_x}{dt} = - \frac{I_\perp}{I_s \omega_s} A \gamma \cos(\gamma t + \phi ) = - A \cos(\gamma t + \phi )\;.
\label{eq:omega_y}
\end{equation}

\noindent Integrating $\omega_x$ and $\omega_y$ we obtain:

\begin{eqnarray}
\eta &=& -\frac{A}{\gamma} \cos(\gamma t + \phi ) + \theta_{x_0}\;,\\
\xi &=& -\frac{A}{\gamma} \sin(\gamma t + \phi ) + \theta_{y_0}\;.	
\label{eq:theta_equations}
\end{eqnarray}

\noindent In the small angle approximation we impose that $A/\gamma \ll 1$.
Such equations reveal that the spin axis rotates around a fixed direction in space. If that direction is along the z-axis then $\theta_{x0}=\theta_{y0}=0$.
Assuming the initial conditions $\eta(t=0)=\theta_0$ and $\xi(t=0)=0$, and assuming that $A/\gamma=\theta_0$ we get:

\begin{eqnarray}\label{eq1}
\eta &=& \theta_0 \cos(\gamma t )\;, \\
\xi &=& \theta_0 \sin(\gamma t )\;.
\label{eq:theta_equations2}
\end{eqnarray}

\noindent The last equations describe the torque-free precession of the spin axis that rotates in space at a fixed angle $\theta_0$ respect to the z-axis with a frequency of the precession motion given by $\gamma = \omega_s I_s/I_\perp$ \citep{kleppner}.\\
Considering a thin disc we get $I_s=2I_\perp$ and so $\gamma=2\omega_s$, thus the disc wobbles twice as fast as it spins.\\

\noindent Finally the apparent rate of a thin disc precession to an observer on the rigid body is:
\begin{equation}
\gamma^\prime = \gamma -\omega_s=\omega_s(\frac{I_s-I_\perp}{I_\perp}) \sim \omega_s
\end{equation}

\section{3D Jones Matrices calculation}
\label{app:stokes_pola}
We detail, in this appendix, the 3D Jones Matrix description of a rotating HWP.
The Jones matrices used here are the ones describing the rotation about the two axes orthogonal to the propagation direction (with their inverse transformations):

\begin{equation}
J_{rotX} (\eta)=
\begin{bmatrix}
1  & 0 & 0\\ 0 & \cos(\eta) & -\sin(\eta)\\
0 & \sin(\eta) & \cos(\eta)\\
\end{bmatrix}, \qquad J_{rotY}(\xi)= 
\begin{bmatrix}
 \cos(\xi)  & 0 & \sin(\xi)\\
 0 & 1 & 0\\
 -\sin(\xi) & 0 & \cos(\xi)\\
 \end{bmatrix}.
 \label{eq:matrix_rotation}
\end{equation}

\noindent The angles $\eta$ and $\xi$ are the precession angles $\eta$ and $\xi$ derived in Section~\ref{sec:precessing_hwp_theory} (Eq.~\ref{eq:theta_x}-\ref{eq:theta_y}).

\noindent The Jones matrix for an half-wave plate with the fast axis at angle $\theta$ with respect to the horizontal axis is:

\begin{equation}
	J_{HWP} (\theta)=
	\begin{bmatrix}
	\cos(2 \theta) & \sin(2 \theta) & 0\\
	\sin(2 \theta) & -\cos(2 \theta) & 0\\
	0  & 0 & 1\\
	\end{bmatrix}
    \label{eq:matrix_hwp}
\end{equation}

In the end, the matrix for a linear polarizer that transmits the horizontal component of a light beam is:
\begin{equation}
	J_{pol}=
	\begin{bmatrix}
	1 & 0 & 0\\
	0 & 0 & 0\\
	0 & 0 & 0\\
	\end{bmatrix}
    \label{eq:matrix_polarizer}
\end{equation}

\subsection{No wobbled case}
Combining such matrices we get a general formula for the Jones vector at the on-axis detector of a polarimeter which modulating element precesses:

\begin{equation}
	i_{out}=J_{pol} \cdot J_{RotY}^{-1}(\xi) \cdot J_{RotX}^{-1}(\eta) \cdot J_{HWP}(\theta) \cdot J_{RotX}(\eta) \cdot J_{RotY}(\xi) \cdot i_{in}
    \label{eq:intensity_product}
\end{equation}

\noindent These calculations have been realized with the Python package \textit{SymPy\footnote{http://www.sympy.org/en/index.html}} which allows symbolic computations with matrices and vectors.

For the ideal case with $\eta=\xi=0^\circ$, when no precession occurs, we get the outgoing Jones vector for an ideal polarimeter:

\begin{equation}
i_{out}=J_{pol} \cdot J_{HWP}(\theta) \cdot i_{in}
=\left[\begin{matrix}E_x \cos{\left (2 \theta \right )} + E_y \sin{\left (2 \theta \right )}\\0\\0\end{matrix}\right]
\label{eq:intensity_vector_matrix}
\end{equation}

\noindent So the Intensity:
\begin{equation}
I= i_{out_x}^2 + i_{out_y}^2 +i_{out_z}^2   = \left(E_x \cos{\left (2 \theta \right )} + E_y \sin{\left (2 \theta \right )}\right)^{2}
\label{eq:intensity_modulated}
\end{equation}

\noindent Where we recognize the modulation terms, at 4 times the frequency of the HWP rotation, within the expressions $\cos^2(2\theta)$, $\sin^2(2\theta)$ and $\cos(2\theta) \sin(2\theta)$. Passing through the Stokes parameters ($T$,$Q$,$U$) we get back the intensity at the same detector for an ideal polarimeter:
\begin{eqnarray}
I&=&\left(E_x \cos{\left (2 \theta \right )} + E_y \sin{\left (2 \theta \right )}\right)\left(E_x^* \cos{\left (2 \theta \right )} + E_y^* \sin{\left (2 \theta \right )}\right)=\nonumber\\
&=&E_x^2\cos^2(2\theta)+E_y^2\sin^2(2\theta) + \sin(2\theta)\cos(2\theta)(E_xE_y^*+E_yE_x^*)=\nonumber\\
&=&\frac{T+Q}{2}\cos^2(2\theta)+\frac{T-Q}{2}\sin^2(2\theta) + \sin(2\theta)\cos(2\theta)\frac{U}{2}=\nonumber\\
&=&\frac{1}{2}\left[T+Q\cos(4\theta)+U\sin(4\theta)\right]
\label{eq:intensity_modulated2},	
\end{eqnarray}

\noindent where the Stokes components $T$,$Q$,$U$ are then defined as usual:

\begin{equation}
T=\langle E_xE_x^*\rangle + \langle E_yE_y^* \rangle   \qquad Q=\langle E_xE_x^*\rangle -\langle E_yE_y^* \rangle \qquad U= \langle E_xE_y^*\rangle + \langle E_yE_x^* \rangle.
\label{eq:stokes_parameters}
\end{equation}

\subsection{No wobbled case}
As for a general polarimeter with a wobbling HWP, we get a general formula for the Jones vector at the on-axis detector:

\begin{equation}
\label{eq:intensity_vector}
	i_{out}=J_{pol} \cdot J_{RotY}^{-1}(\xi) \cdot J_{RotX}^{-1}(\eta) \cdot J_{HWP}(\theta) \cdot J_{RotX}(\eta) \cdot J_{RotY}(\xi) \cdot i_{in}
\end{equation}

For the general case with $\eta$, $\xi \neq 0^\circ$ we find the intensity at the detector: 
\begin{equation}
\label{eq:intensity_general_symbol}
I= i_{out_x}^2 + i_{out_y}^2 +i_{out_z}^2  =\left( E_x g - E_y f\right)^{2}
\end{equation}

\noindent where we define the modulating functions $g(\xi,\theta)$ and 
$f(\eta,\xi,\theta)$: 
\begin{eqnarray}
\label{eq:g_f}
g&=&\sin^{2}{\left (\xi \right )} + \cos^{2}{\left (\xi \right )} \cos{\left (2 \theta \right )}\nonumber ,\\
f&=&2 \sin{\left (\theta \right )} \cos{\left (\xi \right )} \left(\sin{\left (\eta \right )} \sin{\left (\xi \right )} \sin{\left (\theta \right )} -  \cos{\left (\eta \right )} \cos{\left (\theta \right )} \right)
\end{eqnarray}

\noindent We can write the outgoing intensity through the Stokes parameters as follows:
\begin{eqnarray}
\label{eq:intensity_general}
I&=&(E_x g - E_y f)(E_x^* g - E_y^* f) =\nonumber  \\ 
&=&E_x^2 g^2 + E_y^2 f^2 - E_x E_y^* g f - E_y E_x^* g f =\nonumber \\
&=&\frac{T+Q}{2}g^2 + \frac{T-Q}{2}f^2 - U g f =\nonumber \\
&=&\frac{1}{2} \left( g^2 + f^2  \right) T + \frac{1}{2} \left( g^2 - f^2  \right) Q + g f U
\end{eqnarray}

\noindent The last equation gives the intensity at the on-axis detector of a Stokes polarimeter with its modulating element describing a torque-free precession. It is not merely a function of the HWP orientation about the $\hat{z}$ axis, it also depends on the displacements from the $\hat{x}$ and $\hat{y}$ axes due to the precession.

\section{3D M\"uller Matrices calculation}
\label{app:stokes_pola2}
\HW{Since the M\"uller formalism is required to propagate partially polarized light (like che CBM one), we need to calculate the Eq.~\ref{eq:intensity_modulated2} and Eq.~\ref{eq:intensity_general} by using M\"uller formalism. 
3D M\"uller matrices $\textsc{M}_{ij}$ are related to Jones matrices \citep{Samim:article} by:}
\begin{equation}
M_{ij}= tr( \sigma_i \cdot J \cdot \sigma_j \cdot J^{\dagger})  
\label{eq:Muller_to_Jones}
\end{equation}
\HW{where $J$ is the associated Jones matrix, and $\sigma_n$ ($n=[0,...,8]$) are the Gell-Mann matrices. The Eq.~\ref{eq:Muller_to_Jones} is valid if and only if we use trace-normalized Gell-Mann matrices \citep{Gell-Mann:article, Sheppard:16} defined as follows: }
\begin{eqnarray}
\sigma_0 =\frac{1}{\sqrt{3}}\left[\begin{matrix}1 & 0 & 0\\0 & 1 & 0\\0 & 0 & 1\end{matrix}\right], \quad 
\sigma_1 =\frac{1}{\sqrt{2}}\left[\begin{matrix}0 & 1 & 0\\1 & 0 & 0\\0 & 0 & 0\end{matrix}\right], \quad 
\sigma_2 =\frac{1}{\sqrt{2}}\left[\begin{matrix}0 & 0 & 1\\0 & 0 & 0\\1 & 0 & 0\end{matrix}\right],  \quad
\nonumber\\
\sigma_3 =\frac{1}{\sqrt{2}}\left[\begin{matrix}0 & - i & 0\\i & 0 & 0\\0 & 0 & 0\end{matrix}\right], \quad
\sigma_4 =\frac{1}{\sqrt{2}}\left[\begin{matrix}1 & 0 & 0\\0 & -1 & 0\\0 & 0 & 0\end{matrix}\right], \quad 
\sigma_5 =\frac{1}{\sqrt{2}}\left[\begin{matrix}0 & 0 & 0\\0 & 0 & 1\\0 & 1 & 0\end{matrix}\right], \quad 
\nonumber\\
\sigma_6 =\frac{1}{\sqrt{2}}\left[\begin{matrix}0 & 0 & i\\0 & 0 & 0\\-i & 0 & 0\end{matrix}\right], \quad 
\sigma_7 =\frac{1}{\sqrt{2}}\left[\begin{matrix}0 & 0 & 0\\0 & 0 & - i\\0 & i & 0\end{matrix}\right], \quad 
\sigma_8 =\frac{1}{\sqrt{6}}\left[\begin{matrix} 1 & 0 & 0\\0 & 1 & 0\\0 & 0 & - 2 \end{matrix}\right]
\label{eq:gellman}
\end{eqnarray}
\HW{By using the common polarization matrix:}
\begin{equation}
\vec{P} = \langle \vec{E} \vec{E}^* \rangle = \left[\begin{matrix}\langle E_xE_x^* \rangle & \langle E_xE_y^*\rangle & \langle E_xE_z^*\rangle\\
\langle E_yE_x^*\rangle & \langle E_yE_y^*\rangle & \langle E_yE_z^*\rangle\\
\langle E_zE_x^*\rangle & \langle E_zE_y^*\rangle & \langle E_zE_z^*\rangle\end{matrix}\right],
\end{equation}
\HW{we can define the Stokes vector in the 3D formalism}:
\begin{equation}
s=\left[\begin{matrix}\Delta_{0}\\\Delta_{1}\\\Delta_{2}\\\Delta_{3}\\\Delta_{4}\\\Delta_{5}\\\Delta_{6}\\\Delta_{7}\\\Delta_{8}\end{matrix}\right] = 
\left[\begin{matrix} \frac{1}{\sqrt{3}}(\langle E_xE_x^*\rangle + \langle E_yE_y^*\rangle + \langle E_zE_z^*\rangle) \\
\frac{1}{\sqrt{2}}(\langle E_xE_y^*\rangle + \langle E_yE_x^*\rangle) \\
\frac{1}{\sqrt{2}}(\langle E_zE_x^*\rangle + \langle E_xE_z^*\rangle) \\
\frac{i}{\sqrt{2}}(\langle E_xE_y^*\rangle - \langle E_yE_x^*\rangle) \\
\frac{1}{\sqrt{2}}(\langle E_xE_x^*\rangle - \langle E_yE_y^*\rangle) \\
\frac{1}{\sqrt{2}}(\langle E_yE_z^*\rangle + \langle E_zE_y^*\rangle) \\
\frac{i}{\sqrt{2}}(\langle E_zE_x^*\rangle - \langle E_xE_z^*\rangle) \\
\frac{i}{\sqrt{2}}(\langle E_yE_z^*\rangle - \langle E_zE_y^*\rangle) \\
\frac{1}{\sqrt{6}}(\langle E_xE_x^*\rangle + \langle E_yE_y^*\rangle - 2\langle E_zE_z^*\rangle)
\end{matrix}\right]
\label{eq:stokesvector}
\end{equation}

\noindent
\HW{The conventional 2D Stokes parameters are related to the 3D Stokes parameters (optical ordering) by}
\begin{equation}
T = \sqrt{\frac{2}{3}}\left(\Delta_0 +\frac{1}{\sqrt{2}}\Delta_8\right), \quad
Q = \Delta_4, \quad
U = \Delta_1, \quad
V = \Delta_3
\label{eq:stokes_relation}
\end{equation}

\subsection{No wobbled case}
\HW{From Eq.~\ref{eq:Muller_to_Jones} we can find the analogous 3D M\"uller matrices for the Jones HWP matrix (Eq.~\ref{eq:matrix_hwp}) and for the Jones polarizer matrix (Eq.~\ref{eq:matrix_polarizer}):}

\begin{align}
M_{pol x} &= 
\left[\begin{matrix}\frac{1}{3} & 0 & 0 & 0 & \frac{\sqrt{6}}{6} & 0 & 0 & 0 & \frac{\sqrt{2}}{6}\\0 & 0 & 0 & 0 & 0 & 0 & 0 & 0 & 0\\0 & 0 & 0 & 0 & 0 & 0 & 0 & 0 & 0\\0 & 0 & 0 & 0 & 0 & 0 & 0 & 0 & 0\\\frac{\sqrt{6}}{6} & 0 & 0 & 0 & \frac{1}{2} & 0 & 0 & 0 & \frac{\sqrt{3}}{6}\\0 & 0 & 0 & 0 & 0 & 0 & 0 & 0 & 0\\0 & 0 & 0 & 0 & 0 & 0 & 0 & 0 & 0\\0 & 0 & 0 & 0 & 0 & 0 & 0 & 0 & 0\\\frac{\sqrt{2}}{6} & 0 & 0 & 0 & \frac{\sqrt{3}}{6} & 0 & 0 & 0 & \frac{1}{6}\end{matrix}\right] \\
M_{HWP}(\theta) &=
\left[\begin{matrix}1 & 0 & 0 & 0 & 0 & 0 & 0 & 0 & 0\\
0 & - \cos{\left (4 \theta \right )} & 0 & 0 & \sin{\left (4 \theta \right )} & 0 & 0 & 0 & 0\\
0 & 0 & \cos{\left (2 \theta \right )} & 0 & 0 & \sin{\left (2 \theta \right )} & 0 & 0 & 0\\
0 & 0 & 0 & -1 & 0 & 0 & 0 & 0 & 0\\
0 & \sin{\left (4 \theta \right )} & 0 & 0 & \cos{\left (4 \theta \right )} & 0 & 0 & 0 & 0\\
0 & 0 & \sin{\left (2 \theta \right )} & 0 & 0 & - \cos{\left (2 \theta \right )} & 0 & 0 & 0\\
0 & 0 & 0 & 0 & 0 & 0 & \cos{\left (2 \theta \right )} & - \sin{\left (2 \theta \right )} & 0\\
0 & 0 & 0 & 0 & 0 & 0 & - \sin{\left (2 \theta \right )} & - \cos{\left (2 \theta \right )} & 0\\0 & 0 & 0 & 0 & 0 & 0 & 0 & 0 & 1\end{matrix}\right]
\end{align}

\noindent \HW{and by combining the previous matrices we can find the 3D polarimeter Stokes vector ($\eta=\xi=0^\circ$):}
\begin{equation}
s_{out} = M_{pol x} \cdot M_{HWP}(\theta) \cdot s_{in} = M_{pol x} \cdot M_{HWP}(\theta) \cdot \left[\begin{matrix}\Delta_{0}\\\Delta_{1}\\\Delta_{2}\\\Delta_{3}\\\Delta_{4}\\\Delta_{5}\\\Delta_{6}\\\Delta_{7}\\\Delta_{8}\end{matrix}\right]
=
\left[\begin{matrix} \frac{1}{3}\Delta_0 + \frac{\sqrt{6}}{6} \sin{\left (4 \theta \right )}\Delta_1 + \frac{\sqrt{6}}{6}\cos{\left (4 \theta \right )}\Delta_4 + \frac{\sqrt{2}}{6}\Delta_8 \\
0 \\
0 \\
0 \\
\frac{\sqrt{6}}{6}\Delta_0 + \frac{1}{2} \sin{\left (4 \theta \right )}\Delta_1 + \frac{1}{2} \cos{\left (4 \theta \right )}\Delta_4 + \frac{\sqrt{3}}{6}\Delta_8\\
0 \\
0 \\
0 \\
\frac{\sqrt{2}}{6}\Delta_0 + \frac{\sqrt{3}}{6} \sin{\left (4 \theta \right )}\Delta_1 + \frac{\sqrt{3}}{6} \cos{\left (4 \theta \right )}\Delta_4 + \frac{1}{6}\Delta_8
\end{matrix}\right]
\end{equation}

\noindent
\HW{Thanks to Eq.~\ref{eq:stokes_relation} we can derive the intensity:}
\begin{equation}
I = \frac{T}{2} + \frac{ Q}{2} \cos{\left (4 \theta \right )} +  \frac{U}{2} \sin{\left (4 \theta \right )}
\label{eq:stokes_polarimeter_from_Muller}
\end{equation}

\noindent
\HW{Note that the Eq.~\ref{eq:stokes_polarimeter_from_Muller} and Eq.~\ref{eq:intensity_modulated2} give the same result.}

\subsection{Wobbled case}
\HW{For a wobbling HWP we need to calculate the M\"uller rotation matrices (Eq.~\ref{eq:matrix_rotation}) from Eq.~\ref{eq:Muller_to_Jones}:}
\begin{equation}
M_{RotX}(\eta)=\left[\begin{matrix}1 & 0 & 0 & 0 & 0 & 0 & 0 & 0 & 0\\0 & \cos{\left (\eta \right )} & 0 & 0 & 0 & \sin{\left (\eta \right )} & 0 & 0 & 0\\0 & 0 & \cos{\left (2 \eta \right )} & 0 & - \frac{1}{2} \sin{\left (2 \eta \right )} & 0 & 0 & 0 & - \frac{\sqrt{3}}{2} \sin{\left (2 \eta \right )}\\0 & 0 & 0 & \cos{\left (\eta \right )} & 0 & 0 & 0 & - \sin{\left (\eta \right )} & 0\\0 & 0 & \frac{1}{2} \sin{\left (2 \eta \right )} & 0 & \frac{1}{2} \cos^{2}{\left (\eta \right )} + \frac{1}{2} & 0 & 0 & 0 & - \frac{\sqrt{3}}{2} \sin^{2}{\left (\eta \right )}\\0 & - \sin{\left (\eta \right )} & 0 & 0 & 0 & \cos{\left (\eta \right )} & 0 & 0 & 0\\0 & 0 & 0 & 0 & 0 & 0 & 1 & 0 & 0\\0 & 0 & 0 & \sin{\left (\eta \right )} & 0 & 0 & 0 & \cos{\left (\eta \right )} & 0\\0 & 0 & \frac{\sqrt{3}}{2} \sin{\left (2 \eta \right )} & 0 & - \frac{\sqrt{3}}{2} \sin^{2}{\left (\eta \right )} & 0 & 0 & 0 & - \frac{3}{2} \sin^{2}{\left (\eta \right )} + 1\end{matrix}\right]
\end{equation}

\begin{equation}
M_{RotY}(\xi)=\left[\begin{matrix}1 & 0 & 0 & 0 & 0 & 0 & 0 & 0 & 0\\0 & \cos{\left (\xi \right )} & - \sin{\left (\xi \right )} & 0 & 0 & 0 & 0 & 0 & 0\\0 & \sin{\left (\xi \right )} & \cos{\left (\xi \right )} & 0 & 0 & 0 & 0 & 0 & 0\\0 & 0 & 0 & \cos{\left (\xi \right )} & 0 & 0 & - \sin{\left (\xi \right )} & 0 & 0\\0 & 0 & 0 & 0 & \frac{1}{2} \cos^{2}{\left (\xi \right )} + \frac{1}{2} & \frac{1}{2} \sin{\left (2 \xi \right )} & 0 & 0 & \frac{\sqrt{3}}{2} \sin^{2}{\left (\xi \right )}\\0 & 0 & 0 & 0 & - \frac{1}{2} \sin{\left (2 \xi \right )} & \cos{\left (2 \xi \right )} & 0 & 0 & \frac{\sqrt{3}}{2} \sin{\left (2 \xi \right )}\\0 & 0 & 0 & \sin{\left (\xi \right )} & 0 & 0 & \cos{\left (\xi \right )} & 0 & 0\\0 & 0 & 0 & 0 & 0 & 0 & 0 & 1 & 0\\0 & 0 & 0 & 0 & \frac{\sqrt{3}}{2} \sin^{2}{\left (\xi \right )} & - \frac{\sqrt{3}}{2} \sin{\left (2 \xi \right )} & 0 & 0 & - \frac{3}{2} \sin^{2}{\left (\xi \right )} + 1\end{matrix}\right]
\end{equation}

\noindent
\HW{Finally we can calculate the 3D M\"uller matrix for a Stokes polarimeter with a wobbling HWP:}
\begin{equation}
M_{\textit{SP}_\textit{wob}} = M_{polx} \cdot M_{RotY}^{-1}(\xi) \cdot M_{RotX}^{-1}(\eta) \cdot M_{HWP}(\theta) \cdot M_{RotX}(\eta) \cdot M_{RotY}(\xi) = \left[\begin{matrix}
\frac{1}{3} & m_{01} & m_{02} & 0 & m_{04} & m_{05} & 0 & 0 & m_{08}\\
0 & 0 & 0 & 0 & 0 & 0 & 0 & 0 & 0\\
0 & 0 & 0 & 0 & 0 & 0 & 0 & 0 & 0\\
0 & 0 & 0 & 0 & 0 & 0 & 0 & 0 & 0\\
\frac{\sqrt{6}}{6} & \frac{\sqrt{6}}{2}m_{01} & \frac{\sqrt{6}}{2}m_{02} & 0 & \frac{\sqrt{6}}{2}m_{04} & \frac{\sqrt{6}}{2}m_{05} & 0 & 0 & \frac{\sqrt{6}}{2}m_{08}\\
0 & 0 & 0 & 0 & 0 & 0 & 0 & 0 & 0\\
0 & 0 & 0 & 0 & 0 & 0 & 0 & 0 & 0\\
0 & 0 & 0 & 0 & 0 & 0 & 0 & 0 & 0\\
\frac{\sqrt{2}}{6} & \frac{\sqrt{2}}{2}m_{01} & \frac{\sqrt{2}}{2}m_{02} & 0 & \frac{\sqrt{2}}{2}m_{04} & \frac{\sqrt{2}}{2}m_{05} & 0 & 0 & \frac{\sqrt{2}}{2}m_{08}\end{matrix}\right]
\end{equation}

\noindent
\HW{For the sake of clarity we define the following equations we will use from now on: }
\begin{align}
A &= \sin^{2}{\left (\eta \right )} \sin^{2}{\left (\xi \right )} + \sin^{2}{\left (\xi \right )}-1\\
B &=\sin{\left (\eta \right )} \sin{\left (2 \xi \right )}\\
C &=\sin{\left (2 \eta \right )} \sin^{2}{\left (\xi \right )}\\
D &=\sin{\left (2 \xi \right )} \cos{\left (\eta \right )}\\
E &=3 \sin^{2}{\left (\eta \right )} \sin^{2}{\left (\xi \right )} - 3 \sin^{2}{\left (\xi \right )} + 1
\end{align}

\begin{eqnarray}
m_{01}=\frac{\sqrt{6}}{12}(E\sin{\left (2 \eta \right )} \sin{\left (\xi \right )} - 2 \left(A \sin{\left (4 \theta \right )} + B \cos{\left (4 \theta \right )}\right) \cos{\left (\eta \right )} \cos{\left (\xi \right )} - \left(- A \cos{\left (4 \theta \right )} + B \sin{\left (4 \theta \right )}\right) \sin{\left (2 \eta \right )} \sin{\left (\xi \right )} \nonumber\\
- 2 \left(C \sin{\left (2 \theta \right )} + D \cos{\left (2 \theta \right )}\right) \sin{\left (\eta \right )} \cos{\left (\xi \right )} - 2 \left(C \cos{\left (2 \theta \right )} - D \sin{\left (2 \theta \right )}\right) \sin{\left (\xi \right )} \cos{\left (2 \eta \right)} )
\end{eqnarray}

\begin{eqnarray}
m_{02}=
\frac{\sqrt{6}}{12} ( E\sin{\left (2 \xi \right )} \cos^{2}{\left (\eta \right )} + 2\left(A \sin{\left (4 \theta \right )} + B \cos{\left (4 \theta \right )}\right) \sin{\left (\eta \right )} \cos{\left (2 \xi \right )} - \frac{1}{2} \left(A \cos{\left (4 \theta \right )} - B \sin{\left (4 \theta \right )}\right) \left(3 - \cos{\left (2 \eta \right )}\right) \sin{\left (2 \xi \right )} \nonumber\\
- 2 \left(C \sin{\left (2 \theta \right )} + D \cos{\left (2 \theta \right )}\right) \cos{\left (\eta \right )} \cos{\left (2 \xi \right )} + \left(C \cos{\left (2 \theta \right )} - D \sin{\left (2 \theta \right )}\right) \sin{\left (2 \eta \right )} \sin{\left (2 \xi \right )} )
\end{eqnarray}

\begin{eqnarray}
m_{04}=
\frac{\sqrt{6}}{12} (E\left(\cos^{2}{\left (\eta \right )} \cos^{2}{\left (\xi \right )} - 2\cos^{2}{\left (\eta \right )} + 1\right) - \left(A \sin{\left (4 \theta \right )} + B \cos{\left (4 \theta \right )}\right) \sin{\left (\eta \right )} \sin{\left (2 \xi \right )} \nonumber\\
-\left(A \cos{\left (4 \theta \right )} - B \sin{\left (4 \theta \right )}\right) \left(2 \cos^{2}{\left (\eta \right )} + 2 \cos^{2}{\left (\xi \right )} - \cos^{2}{\left (\eta \right )} \cos^{2}{\left (\xi \right )}  - 1 \right) + \nonumber\\
\left(C \sin{\left (2 \theta \right )} + D \cos{\left (2 \theta \right )}\right) \sin{\left (2 \xi \right )} \cos{\left (\eta \right )} + \frac{1}{2} \left(C \cos{\left (2 \theta \right )} - D \sin{\left (2 \theta \right )}\right) \left(\cos{\left (2 \xi \right )} - 3\right) \sin{\left (2 \eta \right )} )
\end{eqnarray}

\begin{eqnarray}
m_{05}=
\frac{\sqrt{6}}{12}(- E \sin{\left (2 \eta \right )} \cos{\left (\xi \right )} - 2 \left(A \sin{\left (4 \theta \right )} + B \cos{\left (4 \theta \right )}\right) \sin{\left (\xi \right )} \cos{\left (\eta \right )} + \left(- A \cos{\left (4 \theta \right )} + B \sin{\left (4 \theta \right )}\right) \sin{\left (2 \eta \right )} \cos{\left (\xi \right )} - \nonumber\\
2 \left(C \sin{\left (2 \theta \right )} + D \cos{\left (2 \theta \right )}\right) \sin{\left (\eta \right )} \sin{\left (\xi \right )} + 2 \left(C \cos{\left (2 \theta \right )} - D \sin{\left (2 \theta \right )}\right) \cos{\left (2 \eta \right )} \cos{\left (\xi \right )} )
\end{eqnarray}

\begin{eqnarray}
m_{08}=
\frac{\sqrt{2}}{4} (E\left(\cos^{2}{\left (\eta \right )} \cos^{2}{\left (\xi \right )} - \frac{1}{3}\right) - \left(A \sin{\left (4 \theta \right )} + B \cos{\left (4 \theta \right )}\right) \sin{\left (\eta \right )} \sin{\left (2 \xi \right )} - \left(A \cos{\left (4 \theta \right )} - B \sin{\left (4 \theta \right )}\right) \left(2(\cos^{2}{\left (\xi \right )} - \cos^{2}{\left (\eta \right )}) - 1 \right) + \nonumber\\
\left(C \sin{\left (2 \theta \right )} + D \cos{\left (2 \theta \right )}\right) \sin{\left (2 \xi \right )} \cos{\left (\eta \right )} + \left(C \cos{\left (2 \theta \right )} - D \sin{\left (2 \theta \right )}\right) \sin{\left (2 \eta \right )} \cos^{2}{\left (\xi \right )} )
\end{eqnarray}

\noindent
\HW{By assuming the field entering the polarimeter has $E_z=0$, this is true only for the on-axis detectors and for all the focal plane assuming a telecentric optic system, we can find the general Stokes vector for a wobbling HWP:}
\begin{equation}
s_{out} = M_{\textit{SP}_{wob}} \cdot s_{in}=M_{\textit{SP}_{wob}} \cdot \left[\begin{matrix}\Delta_{0}\\\Delta_{1}\\0\\0\\\Delta_{4}\\0\\0\\0\\\Delta_{8}\end{matrix}\right]
=
M_{\textit{SP}_{wob}} \cdot \left[\begin{matrix}\Delta_{0}\\\Delta_{1}\\0\\0\\\Delta_{4}\\0\\0\\0\\\frac{1}{\sqrt{2}}\Delta_{0}\end{matrix}\right]
=
\left[\begin{matrix} \frac{1}{3}\Delta_0 + m_{01}\Delta_1 + m_{04}\Delta_4 + m_{08}\frac{1}{\sqrt{2}}\Delta_{0} \\
0 \\
0 \\
0 \\
\frac{\sqrt{6}}{6}\Delta_0 + \frac{\sqrt{6}}{2}\left(m_{01}\Delta_1 + m_{04}\Delta_4 + m_{08}\frac{1}{\sqrt{2}}\Delta_{0}\right)\\
0 \\
0 \\
0 \\
\frac{\sqrt{2}}{6}\Delta_0 + \frac{\sqrt{2}}{2}\left(m_{01}\Delta_1 + m_{04}\Delta_4 + m_{08}\frac{1}{\sqrt{2}}\Delta_{0}\right)
\end{matrix}\right]
\label{eq:stokes_vector_woobled_from_Muller}
\end{equation}
\HW{where we used $\Delta_8=\frac{1}{\sqrt{2}}\Delta_{0}$, and thanks to Eq.~\ref{eq:stokes_relation} the equivalent intensity:}
\begin{equation}
    I= \left(\frac{1}{3}+\frac{\sqrt{2}}{2}m_{08}\right)T+\frac{\sqrt{6}}{2}m_{01}U + \frac{\sqrt{6}}{2}m_{04}Q
    \label{eq:intensity_woobled_from_Muller}
\end{equation}

\HW{Note that the components of the output Stokes vector which are not null (Eq.~\ref{eq:stokes_vector_woobled_from_Muller}) are $\Delta_0'$, $\Delta_4'$, $\Delta_8'$, so from the definition of the Stokes vector (Eq.~\ref{eq:stokesvector}) it is clearly $E_z^{out}=0$. This happens because the polarizing grid doesn't permit $E_z^{out}\neq0$ for on-axis rays. If we lost this assumption (i.e. for an off-axis detector) it is easy to verify that the output Stokes vector became a function of $\Delta_0$, $\Delta_1$, $\Delta_2$, $\Delta_4$, $\Delta_5$, $\Delta_8 $, but the components of the output Stokes vector which are not null are always $\Delta_0'$, $\Delta_4'$, $\Delta_8'$ (Eq.~\ref{eq:stokes_vector_woobled_from_Muller_gen}).}

\begin{equation}
s_{out} = M_{\textit{SP}_{wob}} \cdot s_{in}=
M_{\textit{SP}_{wob}} \cdot \left[\begin{matrix}\Delta_{0}\\\Delta_{1}\\\Delta_{2}\\\Delta_{3}\\\Delta_{4}\\\Delta_{5}\\\Delta_{6}\\\Delta_{7}\\\Delta_{8}\end{matrix}\right]
=
\left[\begin{matrix} \frac{1}{3}\Delta_0 + m_{01}\Delta_1 + m_{02}\Delta_2 + m_{04}\Delta_4 + m_{05}\Delta_5 + m_{08}\Delta_{8} \\
0 \\
0 \\
0 \\
\frac{\sqrt{6}}{6}\Delta_0 + \frac{\sqrt{6}}{2}\left(m_{01}\Delta_1 + m_{02}\Delta_2 + m_{04}\Delta_4 + m_{05}\Delta_5 + m_{08}\Delta_{8}\right)\\
0 \\
0 \\
0 \\
\frac{\sqrt{2}}{6}\Delta_0 + \frac{\sqrt{2}}{2}\left(m_{01}\Delta_1 + m_{02}\Delta_2 + m_{04}\Delta_4 + m_{05}\Delta_5 + m_{08}\Delta_{8}\right)
\end{matrix}\right]
\label{eq:stokes_vector_woobled_from_Muller_gen}
\end{equation}


\bibliographystyle{aa}
\bibliography{aanda}

\end{document}